\documentclass[letterpaper]{JHEP3}
\usepackage{epsfig}
\usepackage{graphicx}
\def\IR{{\hbox{{\rm I}\kern-.2em\hbox{\rm R}}}}
\newcommand{\be}{\begin{equation}}
\newcommand{\ee}{\end{equation}}
\newcommand{\beq}{\begin{equation}}
\newcommand{\eeq}{\end{equation}}
\newcommand{\beqa}{\begin{eqnarray}}
\newcommand{\eeqa}{\end{eqnarray}}

\preprint{{\small hep-th/0303035}} 

\title{SD-brane gravity fields and rolling tachyons}

\author{ Fr\'ed\'eric Leblond\footnote{E-mail: {\tt
fleblond@perimeterinstitute.ca}} \ and Amanda W.
Peet\footnote{E-mail: {\tt peet@physics.utoronto.ca}}\\ $^{*}$
Department of Physics, McGill University, Montr\'eal, Qu\' ebec
H3A 2T8 Canada
\\ $^{*}$ Perimeter Institute for
Theoretical Physics, Waterloo, Ontario N2J 2W9 Canada \\
$^{\dagger}$ Radcliffe Institute, Harvard University, Cambridge,
MA 02138 USA \\ $^{\dagger}$ Department of Physics,
University of Toronto, Toronto, Ontario M5S 1A7 Canada \\
$^{\dagger}$ Cosmology and Gravity Program, Canadian Institute for
Advanced Research}

\date{March, 2003}
\abstract{S(pacelike)D-branes are objects arising naturally in
string theory when Dirichlet boundary conditions are imposed on
the time direction.  SD-brane physics is inherently
time-dependent.  Previous investigations of gravity fields of
SD-branes have yielded undesirable naked spacelike singularities.
We set up the problem of coupling the most relevant open-string
tachyonic mode to massless closed-string modes in the bulk, with
backreaction and Ramond-Ramond fields included.  We find solutions
numerically in a self-consistent approximation; our solutions are
naturally asymptotically flat and time-reversal asymmetric.  We
find completely nonsingular evolution; in particular, the dilaton
and curvature are well-behaved for all time.  The essential
mechanism for spacetime singularity resolution is the inclusion of
full backreaction between the bulk fields and the rolling tachyon.
Our analysis is not the final word on the story, because we have
to make some significant approximations, most notably homogeneity
of the tachyon on the unstable branes.  Nonetheless, we provide
significant progress in plugging a gaping hole in prior
understanding of the gravity fields of SD-branes.}


\begin{document}
\setcounter{footnote}{0}
%====================================================================+
\section{Introduction}\label{section:introduction}

SD-branes in string theory were first studied by Gutperle and
Strominger in ref.~\cite{gutperle}. They were introduced as
objects arising when Dirichlet boundary conditions on open strings
are put on the time coordinate, as well as on spatial coordinates.
SD-branes are not supersymmetric objects, which makes them hard to
handle but potentially very interesting. The boundary conditions
for SD-branes imply that they live for only an ``instant'' of
time, and so the worldvolume is purely spatial. SD-branes should
not be confused with instantons, because they live out their lives
in Lorentzian signature context. Recent discussion of the relation
between SD-branes and instantons may be found in section 6 of
ref.~\cite{andylastweek}.

SD-branes are especially interesting objects to study in the
context of tachyon condensation, which will be the arena of our
investigation. SD-branes are indeed inherently related to the
general study of time dependence in string theory. One of the
original goals of ref.~\cite{gutperle} was in fact to seek
examples of gauge/gravity dualities where a time direction on the
gravity side is holographically reconstructed by a Euclidean field
theory.\footnote{An attempt to find a realization for such a
duality is the dS/CFT correspondence \cite{dscft} -- see also
ref.~\cite{tale} for an extensive list of references.}

There have been several investigations of SD-branes since they
were introduced \cite{gutperle2, KMP, strominger, buchel, roy,
sbranes, andylastweek}.  Most of them involve taking the limit
$g_s\rightarrow 0$, the regime in which perturbative string
computations can be done.  As the tachyon rolls down its potential
hill, there is a divergence in production of higher mass open
string modes \cite{strominger,strominger3,andylastweek}. This
divergence occurs before the tachyon gets to the bottom of the
potential well, as it must because there are no perturbative open
string excitations around the true minimum of the tachyon
potential \cite{vacuum}. Also, the time taken to convert the
energy of the rolling tachyon into these massive open string modes
is of order\footnote{In later sections we will see that our
calculation, which includes gravity backreaction, agrees with this
in the sense that the time it takes for the tachyon to decay only
slightly depends on the particular value of $g_{s}$.}
${\cal{O}}(g_{s}^{0})$ .  This analysis was done for a single
SD-brane using CFT methods; analysis for production of massive
closed string modes was also done \cite{OS,maldacena}.

One aspect of SD-brane physics has become clear: that the
decoupling limit applied to SD-branes is not a smooth limit like
it is for regular D-branes. In particular, as $g_s\rightarrow 0$
the brane tachyon becomes decoupled from bulk modes, which were
however the most natural modes into which the initial energy of
the unstable brane should decay.  Then, the endpoint of the
rolling tachyon must include a somewhat mysterious substance
called ``tachyon matter'' \cite{sen2}. Consideration of the full
problem with $g_s$ finite would presumably eliminate the need for
mysterious tachyon matter; this was in fact part of our motivation
for this work.  Regarding production of closed string massive
modes, at very small $g_s$, it seemed that there was some debate
\cite{OS,maldacena} about the form of a divergence.
The results from ref.~\cite{maldacena} make clear that the
divergence depends on the number of dimensions transverse to the
decaying unstable brane: for unstable D$p$-branes with $p<2$ there
was a need to invoke a cutoff to get a finite result.
In any case, unstable brane decay should presumably be a physically
smooth process for $\{g_s,\ell_s\}$ finite.

The point of view that we will be taking is to consider a system of
$N$ SD-branes, with $g_sN$ large, study the overall centre-of-mass
tachyon, and couple it to bulk massless closed string modes.
Obviously, it would be nice to understand the full problem including
coupling to all massive open and closed string modes, but this is a
hard problem beyond our reach.  We will make a beginning here with a
quantitative analysis involving only the lowest modes in each of the
open and closed string sectors.  We believe that our approach, while
``lowbrow'' by comparison to SFT computations, already shows some very
interesting physics.

The punchline of our paper will be this: we find nonsingular solutions
for the evolution of the open string tachyon coupled to bulk
supergravity modes.  This plugs a gaping hole in our previous
understanding \cite{gutperle,gutperle2,KMP} of the supergravity fields
arising from a large number of SD-branes.  All previous attempts at
describing SD-branes in the context of supergravity had found that the
corresponding solutions were plagued with naked spacelike
singularities.  We find that resolution of these singularities is
achieved in a conceptually simple way: by including full backreaction
on the rolling tachyon.

Our investigation can also be considered to shed light on the
question of tachyon cosmology \cite{gibbons} including
Ramond-Ramond fields, the effect of which was ignored in previous
investigations.  Tachyon cosmology itself may not yet provide
realistic models for inflation, nonetheless, see recent work
including, {\it e.g.}, refs.~\cite{cosmotachyonK,cosmotachyon}.
One of the reasons is that, in the low-energy actions used to
describe tachyon cosmology dynamics, there is only one length
scale --- the string length.  It would certainly be interesting if
a mechanism generating a lower scale for inflation were found
within this context. Also, the behavior of inhomogeneities during
the later stage of the roll of the tachyon may be a general
problem \cite{cosmotachyonK}.
Tachyon cosmology involving brane-antibrane annihilation may be
relevant only to a pre-inflationary period, but it is interesting to
analyze the dynamics from the ``top-down'' perspective in string
theory.

The plan of our paper is as follows. We begin by reviewing in
section \ref{section:review} the previous work on gravity fields
of SD-branes, and commenting on the nature of singularities found
in the past.  In section \ref{subsection:ssss} we discuss
pathologies of the solutions found in refs.~\cite{gutperle2,KMP},
where only supergravity fields were considered.  In section
\ref{subsection:buchel} we move to discussing ref.~\cite{buchel},
in which an unstable brane probe was coupled to a $d=4$ SD0-brane
supergravity background; we generalized their arguments but still
find generic singularities in the probe approximation (exceptions
are considered in Appendix~\ref{regular}).
In section \ref{section:newsetup} we set up the full backreacted
problem of interest.  The equations are naturally highly nonlinear,
and since backreaction is essential we have no desire to ignore it or
treat it perturbatively.  We need to use a particular homogeneous
ansatz to facilitate solution of the equations of motion, and we
discuss implications of the ansatz.
In section \ref{section:numeric} we demonstrate numerical solutions of
our backreaction-inclusive equations, and interpret qualitative
features found.  In particular, we follow carefully the evolution of
both the dilaton and curvature invariants.
Lastly, in the Discussion section we summarize our conclusions, open
issues, and directions of future work.

%====================================================================+
\section{Singular supergravity SD-branes: Review}\label{section:review}

In the paper \cite{gutperle}, a small number of supergravity
solutions, thought to be appropriate for a large number $N$ of
SD-branes, were presented. Subsequently, the class of supergravity
solutions was widened considerably, simultaneously by two groups
(refs.~\cite{gutperle2} and \cite{KMP}).  A later paper showed that
these two sets of solutions were equivalent \cite{roy}, by matching
boundary conditions on asymptotic fields and finding the coordinate
transformation explicitly.

%--------------------------------------------------------------------+
\subsection{Sourceless SD-brane supergravity
solutions}\label{subsection:ssss}

The convention of ref.~\cite{gutperle}, which we will use, is that
SD$p$-branes have $(p{+}1)$ worldvolume coordinates.  We call
these ${\vec{y}}$. There is also the time coordinate $t$, and the
$(8{-}p)$ overall transverse coordinates ${\vec{x}}$.

The most general SD$p$-brane supergravity solutions of
refs.~\cite{gutperle2} and \cite{KMP} can then be written in
string frame as follows,
\begin{eqnarray}
\label{bigmetric} ds^{2}_{{\scriptscriptstyle
Sp}}&=&F(t)^{1/2}\beta(t)^G\alpha(t)^H
\left(-dt^2+t^2 dH^{2}_{8-p}({\vec{x}})\right) \nonumber\\
&& +F(t)^{-1/2}\left[\sum_{i=2}^{p+1}
\left({\beta(t)\over\alpha(t)}\right)^{-k_i}(dy^i)^2
+\left({\beta(t)\over\alpha(t)}\right)^{k_{1}+{\tilde{k}}}
(dy^{1})^2\right]\, , \nonumber\\
e^{2\Phi} &=& F(t)^{3-p\over2}
\left({\beta(t)\over\alpha(t)}\right)^{-\sum_{i=2}^{p+1}k_i}\ ,
\label{solupb}\\
C^{(p+1)}&=& \sin\theta\cos\theta\,{C(t)\over F(t)}\, dy^{1}\wedge
\cdots\wedge dy^{p+1}\ ,\nonumber
\end{eqnarray}
where
\begin{equation}
C(t) \equiv \left({\beta(t)\over\alpha(t)}\right)^{k_1}
-\left({\beta(t)\over\alpha(t)}\right)^{{\tilde{k}}}\,,\qquad
F(t)\equiv
\cos^2\theta\left({\beta(t)\over\alpha(t)}\right)^{{\tilde{k}}}
+\sin^2\theta\left({\beta(t)\over\alpha(t)}\right)^{k_1} \ ,
\label{useful}
\end{equation}
and
\begin{equation}\label{alphabeta}
\alpha(t)\equiv 1+\left({\frac{\omega}{t}} \right)^{7-p}\,,\quad
\beta(t)\equiv 1-\left({\frac{\omega}{t}} \right)^{7-p}\,.
\end{equation}
The supergravity equations of motion will be satisfied when the
exponents satisfy the constraints,
\begin{eqnarray}
{\tilde{k}}^2+\sum_{i=1}^{p+1}k_i^2+{7-p\over4}(H-G)^2-4
{8-p\over 7-p}&=&0\ , \nonumber\\
{\tilde{k}}+\sum_{i=1}^{p+1}k_i-{7-p\over2}(H-G)&=&0\ ,
\label{solve}\\
H+G-{4\over 7-p}&=&0\ .  \nonumber
\end{eqnarray}

The general metric above is not isotropic in the worldvolume
directions ${\vec{y}}$.  However, from a microscopic point of
view, one expects that the supergravity solution would have an
isotropic worldvolume. Isotropy in the worldvolume will be
restored in the above solution for the choice
$-k_2=\cdots=-k_{p+1}=k_{1}+{\tilde{k}}\equiv n$.  Unfortunately,
the isotropy requirement excludes SD-brane solutions with regular
Cauchy horizon. In fact, the curvature invariants associated to
all isotropic solutions diverge at $t=\pm \omega$ and $t=0$.
Consequently, although they possess the right symmetries and
``charge'', these solutions do not appear to be well-defined.

Besides, SD-branes should be represented by solutions with,
roughly, three distinct regions: the infinite past with incoming
radiation only in the form of massless closed strings, an
intermediate region with both open and closed strings, and finally
the infinite future with dissipating outgoing radiation in the
form of massless closed strings. But the supergravity solutions of
refs.~\cite{gutperle,gutperle2,KMP} cannot represent this process,
because there are no rules for deciding how to go through the
singular regions (see, however, Appendix~\ref{regular} where we
show that some anisotropic solutions avoid the pathology).

Nevertheless, the isotropic solutions have a positive feature
worth noting: they have the correct asymptotics at large time. In
the limit that the functions $\alpha(t)$ and $\beta(t)$ become
trivial, part of the metric is simply the Milne universe: flat
Minkowski space foliated by hyperbolic sections, \beq
\lim_{t\rightarrow \pm \infty}ds^{2}_{{\scriptscriptstyle Sp}}=
-dt^2+t^2 dH^2_{8-p}({\vec x}) + \sum_{i=1}^{p+2} (d{y}^i)^2 .
\eeq On the other hand, there are (at least) two reasons to
suspect that the above solutions are not the final word in the
SD-brane supergravity story. The first is that there are one too
many parameters in the solution, by comparison to expectations
from microscopics of SD-branes \cite{KMP}.  A possible explanation
for this may be as follows.  In the rest of our paper, we will be
showing that the full coupling/backreaction between the
open-string tachyon and the closed-string bulk modes is crucial
for resolution of spacetime singularities.  It is possible that
the freedom in the supergravity solutions may correspond to a
freedom in picking boundary conditions for the rolling tachyon ---
the coupling to which was not included in
refs.~\cite{gutperle,gutperle2,KMP}.

The most noticeable negative feature of the above solutions is
that the isotropic solutions are nakedly singular.  Quite
generally, nakedly singular spacetimes arising in low-energy
string theory come under immediate suspicion, even though they are
solutions to the supergravity field equations.  No-hair theorems
are usually what we rely on in order to be sure that we have the
unique supergravity solution, but no-hair theorems are never valid
for solutions with {\em naked} spacetime singularities.  It is
worth noting that it has been shown with an explicit
counterexample \cite{emparan} that even no-hair theorems
themselves fail for black holes in $d=5$ with mass and angular
momentum --- and hence the idea of no-hair theorems in all higher
dimensions is under suspicion.  (Nonetheless, with particular
assumptions about field couplings, no-hair theorems can be proven
for static asymptotically flat dilaton black holes
\cite{gibbonshair}. Also, uniqueness of the {\em supersymmetric}
rotating BMPV \cite{BMPV} black hole in $d=5$ has been proven
\cite{reallbmpv}.)  Even if a no-hair theorem appropriate to the
supergravity theory involving SD-branes could be proven, however,
the above solutions we have reviewed would be ruled out as
candidates because their singularities are uncontrollably nasty.
So we have to look elsewhere.

Let us take a brief sidetrip here to comment on the singularity
story for supergravity fields of $N{\gg}1$ regular D-branes with
timelike worldvolume.  Certainly, the geometry of BPS D3-branes is
nonsingular, and there are several other pretty situations known
in the literature where branes ``melt'' into fluxes: the sources
are no longer needed. However, the disappearance of D-brane
sources for supergravity fields only occurs when the branes are
BPS.  If any energy density above BPS is added to these systems,
singularities reappear: this certainly happens for the D3-brane
system.  Also, in the low-energy approximation to string theory,
it is misleading to think of supergravity fields of D$p$-branes as
simple condensates of massless closed string modes.  The reason is
directly analogous to the fact that the Coulomb field of an
electron cannot be a photon condensate because the photon is
transverse.  Similarly, Coulomb Ramond-Ramond fields of
D$p$-branes cannot be represented by supergravity fields
alone.\footnote{This is the case even though D$p$-branes are
``solitonic'' in string theory while electrons are fundamental in
QED. The straightforward argument we use here depends only on the
couplings of the charge-carrying objects to the bulk gauge fields.
We thank Abhay Ashtekar for a discussion on this issue.}  This
river runs deeper: in the decoupling limit, resolution of dilaton
and curvature singularities for $p\not =3$ D$p$-branes is in fact
{\em provided} by the gauge theory on the D-branes \cite{IMSY,
awpconfproc}.

Let us now get back to our SD-branes.  The supergravity situation
looks similar to that for non-BPS (ordinary) D-branes: it seems that
brane modes will be required for singularity resolution.  Therefore,
we are motivated to try to solve all problems with prior candidate
SD-brane spacetimes by solving the coupled system of brane tachyon
plus bulk supergravity fields with full backreaction.

%--------------------------------------------------------------------+
\subsection{Unstable brane probes in sourceless SD-brane
backgrounds}\label{subsection:buchel}

The first progress towards the goal of singularity resolution in
SD-brane systems was made by Buchel, Langfelder and Walcher
\cite{buchel}.
% type %
We now briefly review what is, for our purposes, the most relevant
point of their work.

Essentially, they take the reasonable point of view that the
process of creation and subsequent decay of a SD$p$-brane must be
driven by a single open string mode: the tachyon field denoted
$T(t)$, which lives on the associated unstable D$(p{+}1)$-brane.
They use the $p=0$ {\em non}-dilatonic version of the worldvolume
% type %
action \cite{add1,add2,add3,add4,sen2,sen1}
\beq \label{sourcebuc}
S_{brane} = -T_{p+1} \int d^{p+2}y \, e^{-\Phi} V(T) \sqrt{-det \,
\left( {\cal P
}G_{\alpha\beta}+\partial_{\alpha}T\partial_{\beta}T \right)} +
\mu_{p+1}\int f(T)dT\wedge C^{(p+1)}, \eeq
to study the dynamics of the tachyon and its couplings with bulk
(closed string) modes.  The operation ${\cal P}$ is for pullback, and
$\alpha,\beta=1\ldots(p{+}1)$. In section \ref{section:prelim} we will
comment both on the validity of this type of effective action, and on
the expected form for the tachyon potential $V(T)$ and the
Ramond-Ramond coupling $f(T)$.  For now, we just use it.

The way we look at the calculation of ref.~\cite{buchel} is as
follows. Supergravity SD-brane fields should be regarded as
arising directly from a large number of unstable branes. Then,
using the intuition gained from studying the enhan{\c{c}}on
mechanism \cite{enhanconjpp}, it is natural to use an unstable
brane probe to study more substantively the candidate supergravity
solutions of refs.~\cite{gutperle2,KMP}.  Then, we look for
problems arising in the probe calculation.  The idea is that
whenever the probe analysis goes wrong, it signals a pathology for
the gravitational background. There are at least two ways the
probe analysis can signal a problem: infinite energy or pressure
density for the tachyon may be induced, ($\rho_{probe}, p_{probe}
\rightarrow \pm \infty$), or there might not exist any reasonable
solutions for $T(t)$ across the horizon.

So let us consider inserting an unstable brane probe in a
background with fields corresponding to the sourceless SD-brane
supergravity fields, eqs.~(\ref{bigmetric}), of the previous
subsection.  The equation of motion for the open string tachyon
is, generally,
\begin{eqnarray}\label{eomt_one}
&& (-g_{tt})^{\frac{1}{2}} (g_{y^{i}y^{i}})^{\frac{p}{2}}
(g_{y^{1}y^{1}})^{\frac{1}{2}} e^{-\Phi}
\frac{\partial V(T)}{\partial
T}\Delta^{\frac{1}{2}}  \nonumber\\
+ && f(T)\, F^{(p+2)} + \frac{d}{dt}\left( \dot{T}\, V(T)
\frac{(g_{y^{1}y^{1}})^{\frac{1}{2}}(g_{y^{i}y^{i}})^{\frac{p}{2}}}
{e^{\Phi}\Delta^{\frac{1}{2}} (-g_{tt})^{\frac{1}{2}}}\right)=0,
\end{eqnarray}
where ${\dot{\ }}\equiv d/dt$, our notation for the Ramond-Ramond
field strength is $F^{(p+2)}=dC^{(p+1)}$, we have factored out
$T_{p+1}$ by including a factor of $g_s$ in $C^{(p+1)}$, and we
also defined the following expression, \beq \Delta = 1 +
\frac{(\dot{T})^{2}}{g_{tt}}.\eeq

The question needing attention here is whether or not the tachyon
field, regarded as a probe,\footnote{The unstable brane will be a
probe as long as its backreaction is small and can therefore be
treated self-consistently as a perturbation.} is well-behaved when
inserted in the candidate supergravity backgounds
(\ref{bigmetric}). Ref.~\cite{buchel} provided a clear answer for
the case of a $d=4$ SD0-brane with dilaton field set to
zero.\footnote{The authors of ref.~\cite{buchel} also investigated
the effect of tachyon backreaction, but the ansatz they used for
the supergravity fields was not general enough to handle our cases
of interest.  In particular, our general equations {\em do not}
reduce to theirs upon consistent truncation. Also, their
exposition of their backreaction analysis was {\em extremely}
brief.}
Let us now see how this goes.

The $d=4$ SD0-brane background introduced in ref.~\cite{gutperle}
has the form, \beq g_{tt} = -g^{y^{1}y^{1}} =
-\frac{Q^{2}}{\omega^{2}} \frac{t^{2}}{t^{2}-\omega^{2}}, \;\;\;\;
g_{y^{i}y^{i}}=0, \;\;\;\; g_{xx} = \frac{Q^{2}}{\omega^{2}}t^{2},
\eeq and \beq F_{2} = Q \epsilon_{2} \,,\quad \Phi=0. \eeq This
spacetime metric has a regular horizon (a coordinate singularity)
at $t=\omega$, and a genuine timelike curvature singularity at
$t=0$. The expressions for the energy density ($T^{t}_{t}$) and
the pressure ($T^{y^{1}}_{y^{1}}$) associated with the probe are
respectively, \beq \rho_{probe} \sim
\frac{V(T)}{\Delta^{1/2}}\,,\quad p_{probe} \sim V(T)
\Delta^{1/2}. \eeq

The only time (apart from $t=0$) when the probe limit becomes
ill-defined is around $t=\omega$. In fact, in the near horizon
limit, the dynamical quantity $\Delta$ satisfies the simple
ordinary differential equation \beq \Delta^{2} - \Delta +
(t-\omega)\dot{\Delta} = 0. \eeq This has the general solution
\beq \Delta = \frac{t-\omega}{t-\omega + g}, \eeq where $g$ is a
constant of integration. There are two possible solutions at
$t=\omega$: $\Delta=0$ ($g\neq 0$) or $\Delta =1$ ($g=0$).
Clearly, the case for which $\Delta =0$ corresponds to the probe
limit breaking down since the energy density of the unstable brane
diverges. The other possibility, $\Delta=1$, implies that the
time-derivative of the tachyon diverges on the horizon. This last
case is clearly pathological and cannot correspond to a physically
relevant tachyon field solution.

The conclusion is that unstable brane {\em probes} are not
well-defined in the SD0-brane background. Not only are they useless to
resolve the timelike singularity at $t=0$ but, worse, they appear to
generically induce a spacelike curvature singularity on the horizon at
$t=\omega$.  That is, if we take the probe story to be a good
indicator of the story for the full backreacted problem.

One of our first motivations for the work leading to our paper was
to plumb how restricted the conclusions of ref.~\cite{buchel}
were. Did this above story work only for bulk couplings of the
kind arising for SD0-branes in $d=4$, where no dilaton field
appears? Was it true only for the case of SD0-branes, which are a
special case for SD$p$-branes since there can be no anistropy in a
one-dimensional worldvolume?  Are all timelike clothed
singularities turned into naked spacelike singularities by probes?
Could we even trust the probe approximation to tell us anything
about the solution with full backreaction?

The first generalization we considered was to look at an unstable
brane probe in the background of the isotropic SD$p$-brane
solutions of ref.~\cite{KMP}.  However, what we saw there was that
the naked spacelike singularities remained naked spacelike
singularities; the small effect of the probe could not undo that
pathology.  Next, we moved to analyzing anisotropic solutions of
the form (\ref{bigmetric}), those with regular horizons.  Some of
these are actually completely nonsingular; we analyzed the details
of the probe computation in those backgrounds, and the specifics
are recorded in Appendix \ref{regular}.  The results there are
simple to summarize: the solutions with singularities hidden
behind horizons do not give rise to conclusions qualitatively
different than what we have reviewed here for $d=4$ SD0-branes.
The picture therefore remains unsatisfactory.

The upshot, then, is that the probe story does not resolve
singularities found for sourceless supergravity SD-brane solutions.
So we now move to the full backreacted problem for SD$p$-branes in
$d{=}10$, which is the main content of our paper.

%====================================================================+
\section{Supergravity SD-branes with a tachyon
source}\label{section:newsetup}

In this section we find, in the context of supergravity, the
equations of motion associated with the real-time (formation and)
decay of a clump of unstable D-branes.  We begin this section by
writing the form of the action which we will use in our analysis.
We will concentrate on only the most relevant\footnote{Relevant in
the technical sense.} modes in both the open and closed string
sectors; in other words, we keep in our analysis only the
(homogeneous) tachyon and massless bulk supergravity fields.
Potentially, the effect of massive modes could be encoded in a
modified equation of state for the tachyon fluid on the unstable
D-brane.\footnote{We thank Andy Strominger for this suggestion.}
We leave for future work the issue of non-homogeneous tachyonic
modes, and of massive string modes in both the open and closed
string sectors, for the coupled bulk-brane system with full
backreaction.  In order to use the supergravity approximation here
self-consistently, we will take $g_s$ small but $g_s N$ large, and
time-derivatives will be small compared to $\ell_s$.  We will see
that it is simple to choose boundary conditions in our numerical
integration such that these remain true for all time.

%--------------------------------------------------------------------+
\subsection{Preliminaries: action and equations of motion}
\label{section:prelim}

For this section we will be able to suppress R-R Chern-Simons
terms in writing the bulk action.  This is a consistent
truncation, to set the NS-NS two-form to zero throughout the
evolution of the system of interest, as long as consistency
conditions on the R-R fields are satisfied. {\it E.g.} for the
SD2-brane system with R-R field $C^{(3)}$ activated, it is
necessary to make certain that $dC^{(3)}\wedge dC^{(3)}=0$ in
order not to activate the NS-NS two-form and the accompanying
Chern-Simons terms.  Other cases are related to this one by
T-duality.  Therefore, we allow only electric-type coupling of the
SDp-brane to $C^{(p+1)}$ (or equivalently magnetic-type coupling
to $C^{(7-p)}$).  Later we will show that this ansatz is
physically consistent provided we assume that there is $ISO(p+1)$
symmetry along the worldvolume of the SDp-brane, the object we are
interested in.  This is equivalent to considering {\em only the
lowest-mass tachyon}, {\it i.e.}, not allowing any excitations of
the brane tachyon along the spatial worldvolume directions.  Of
course, the R-R field strengths are then very simple:
$G^{(p+2)}=dC^{(p+1)}$, and the string frame bulk action takes the
form \cite{polchinski}
\begin{equation}\label{bulk_action}
S_{{\rm{bulk}}}={\frac{1}{1 6\pi G_{1 0}}} \int d^{1
0}x\,\sqrt{-g}\left\{ e^{-2\Phi}\left[{\cal
R}+4(\partial\Phi)^{2}\right]
-{\frac{1}{2}}\left|dC^{p+1}\right|^{2} \right\} \, ,
\end{equation}
where ${\cal R}$ is the Ricci scalar. We use a mostly plus
signature. In the above conventions, the R-R field solutions
automatically get a factor of $g_s$ (as we mentioned also in the
previous subsection), and
\begin{equation}\label{Newton}
16\pi G_{10}= (2\pi)^{7}g_{s}^{2}\ell_{s}^{8}\,,\quad
\tau_{Dp}={\frac{1}{g_{s}(2\pi)^{p}\ell_{s}^{p+1}}}.
\end{equation}

Our analytical and numerical results in following sections will be
given in {\em string frame}.  If desired, it is easy to convert to
Einstein frame -- with metric ${\tilde{g}}_{\mu\nu}$ -- with
canonical normalization of the metric and positive dilaton kinetic
energy, by using the standard $d=10$ conformal transformation
\begin{equation}\label{string_Einstein}
{\tilde{g}}_{\mu\nu} = e^{-(\Phi-\Phi_\infty)/2}g_{\mu\nu}.
\end{equation}
Stress-tensors are defined in Einstein frame,
\begin{equation}\label{stress_definition}
{\tilde{T}}_{\mu\nu}\equiv{\frac{-1}{\sqrt{-{\tilde{g}}}}}
{\frac{\delta S_{{\rm{matter}}}}{\delta {\tilde{g}}^{\mu\nu}}},
\end{equation}
with the usual
\begin{eqnarray}\label{stress_tensor_CT}
{\tilde{T}}_{\mu\nu}\left[\Phi\right]&=&{\frac{1}{2}}\left[
\partial_{\mu}\Phi\partial_{\nu}\Phi
-{\frac{1}{2}}{\tilde{g}}_{\mu\nu}
({\tilde{\partial\Phi}})^{2}\right]\, , \\
{\tilde{T}}_{\mu\nu}\left[C^{p+1}\right] &=& {\frac{1}{2(p+1)!}}
e^{(3-p)\Phi/2} \left[ {\tilde{G}}_{\mu}^{\
\lambda_{2}\ldots\lambda_{p+2}
}G_{\nu\lambda_{2}\ldots\lambda_{p+2}} - {\frac{1}{2(p+2)}}
{\tilde{g}}_{\mu\nu}{\tilde{G}}^{2} \right].
\end{eqnarray}
We can transform to string-frame ``Einstein'' equation using
standard formul\ae\
\begin{eqnarray}\label{Einstein_string}
&& {\tilde{{\cal R}}}_{\mu\nu} -
{\frac{1}{2}}{\tilde{g}}_{\mu\nu}{\tilde{{\cal R}}}
-{\frac{1}{2}}\left[\partial_{\mu}\Phi\partial_{\nu}\Phi
-{\frac{1}{2}}{\tilde{g}}_{\mu\nu}({\tilde{\partial\Phi}})^{2}
\right] \\
= &&{\cal R}_{\mu\nu} - {\frac{1}{2}}g_{\mu\nu}{\cal R} +
2\left[\nabla_{\mu}\partial_{\nu}\Phi - g_{\mu\nu}\nabla^{2}\Phi +
g_{\mu\nu}(\partial\Phi)^{2}\right]. \nonumber
\end{eqnarray}
For all matter fields except the dilaton, it is therefore obvious
that string frame ``stress-tensors'' take the form
\begin{equation}\label{peasy_string_stress}
T_{\mu\nu} = {\frac{-1}{\sqrt{-g}}} {\frac{\delta S_{\rm
matter}}{\delta g^{\mu\nu}}}.
\end{equation}
For the dilaton we find
\begin{equation}\label{stress_string_dilaton}
T^{\mu}_{\ \nu}\left[\Phi \right] =
2\left[-\nabla^{\mu}\partial_{\nu}\Phi + g^{\mu}_{\ \nu}
\nabla^{2}\Phi-g^{\mu}_{\ \nu}(\partial\Phi)^{2} \right] \, ,
\end{equation}
and, obviously, familiar energy conditions for bulk fields are only
satisfied in Einstein frame, not string frame.

For the bulk field equations, we must include coupling to the brane
tachyon --- this is of course an important point of our paper. Hence,
we now turn to the brane action. The brane theory is that appropriate
to unstable D($p+1$)-branes, for SD$p$-branes. We consider $N$ branes.
At low energy, the overall $U(1)$ center-of-mass tachyon $T$ couples
as follows:
\begin{equation}\label{brane_worldvolume_action}
S_{{\rm{brane}}} = {\frac{\Lambda}{16\pi G_{10}}} \left\{
 \int dt d^{p+1}y \left[ - \ e^{-\Phi} \sqrt{-A}V(T)\right]
+  \int f(T)dT\wedge C^{p+1} \right\} \, ,
\end{equation}
where the matrix $A_{\alpha\beta}$ is defined as
\begin{equation}\label{A_definition}
A_{\alpha\beta} = {\cal{P}}\left(G_{\alpha\beta}+B_{\alpha\beta}
\right) +F_{\alpha\beta} +
\partial_{\alpha}T\partial_{\beta}T \, ,
\end{equation}
where ${\cal{P}}$ stands for pullback.  For the constants, our
conventions are that  $T$ is normalized like $F_{\alpha\beta}$,
and also
\begin{equation}\label{constant_convention}
\Lambda\equiv {\frac{N\mu_{p+1}}{g_{s}}} (16\pi G_{10}) =
(Ng_{s})(2\pi\ell_s)^{6-p}.
\end{equation}
Notice that $\Lambda$ is proportional to $g_{s}N$.  This will be
the {\em sole} continuous\footnote{$\Lambda$ is effectively
continuous in the supergravity approximation since $g_{s}N$ is
large and all derivatives small in string ($\ell_s$) units.}
control parameter associated with the physics of our final
solutions for the coupled tachyon-supergravity system.

It is important to know when we can expect to trust the action we
use. Our approach consists in assuming that the kinetic terms of the
open string tachyon field are re-summed to take a Born-Infeld--like
form. Strictly speaking, this has only be shown to be a valid claim
late in the tachyon evolution. We refer the reader to ref.~\cite{sen1}
for more details on the limits in which
this approximation holds (see also ref.~\cite{kutasov2}).
The functions $V(T)$ and $f(T)$ are therefore not known exactly at
all times. For definiteness in our numerical analysis, we will
choose a specific form and assume that the dynamics of the tachyon
is governed by eq.~(\ref{brane_worldvolume_action}). For a
SD-brane, we make the choice $V(T) = 1/\cosh(T/\sqrt{2})=|f(T)|$
which has been shown to be the correct large-$T$ behavior of the
couplings.  Our results turn out to be quite robust, in that their
qualitative features do not depend on the precise form of $V(T)$
and $f(T)$.

For the remainder of our discussion it will be convenient to use
static gauge, which is an appropriate gauge choice for our problem
of interest.  Therefore, in the following, we will be rather
cavalier about dropping pullback signs.

When we get to solving the coupled brane-bulk equations, it will
be convenient to allow for a density of branes, denoted
$\rho_{\perp}$, in the transverse space:
\begin{equation}\label{branedensity}
\int_{{\rm{brane}}} dt d^{p+1}y \longrightarrow \int_{{\rm{bulk}}}
dt d^{p+1}x d^{8-p}x \rho_{\perp} \, .
\end{equation}
Therefore our brane action becomes
\begin{equation}\label{brane_action}
S_{{\rm{brane}}} = {\frac{\Lambda}{16\pi G_{10}}} \int d^{10}x
\rho_{\perp} \left\{ - V(T)\sqrt{-A}e^{-\Phi} + f(T)
\epsilon^{\lambda_{1}\ldots\lambda_{p+2}}
C_{[\lambda_{2}\ldots\lambda{p+2}} \partial_{\lambda_{1}]}T
\right\},
\end{equation}
where $\epsilon$ is the worldvolume permutation tensor with values
$(0,\pm 1)$.

We are now ready to write down the coupled field equations.  The
simplest bulk equation to pick off is the dilaton.  In string frame we
see immediately that
\begin{equation}\label{bulk_dilaton}
\nabla^{2}\Phi - (\partial\Phi)^{2} + {\frac{1}{4}} {\cal R} =
{\frac{1}{8}} \left({\frac{\Lambda\rho_{\perp}}{\sqrt{-g}}}\right)
e^{+\Phi}V(T)\sqrt{-A},
\end{equation}
and for the Ramond-Ramond field
\begin{equation}\label{bulk_Ramond-Ramond}
\nabla_{\mu}G^{\mu\lambda_{2}\ldots\lambda_{p+2}} =
-\left({\frac{\Lambda\rho_{\perp}}{\sqrt{-g}}}\right) f(T)
\epsilon^{\mu\lambda_{2}\ldots\lambda_{p+2}}\partial_{\mu}T.
\end{equation}
For the metric equation of motion (string frame ``Einstein''
equations), it is convenient to define
\begin{equation}\label{new_einstein}
{\cal R}^{\mu}_{\ \nu} = {\hat{T}}^{\mu}_{\ \nu} \equiv T^{\mu}_{\
\nu} - {\frac{1}{8}} T^\lambda_\lambda .
\end{equation}
Therefore, we have
\begin{equation}\label{hat_dilaton_stress}
{\hat{T}}^{\mu}_{\ \nu} [\Phi]= -2\nabla^{\mu} \partial_{\nu} \Phi
-{\frac{1}{4}} g^{\mu}_{\ \nu} \nabla^2\Phi + {\frac{1}{2}}
(\partial\Phi)^2 g^{\mu}_{\ \nu} .
\end{equation}
For the brane stress-tensor, we need to figure out the dependence
of $\sqrt{-A}$ on $g_{\mu\nu} $ (the Wess-Zumino term clearly does
not contribute). We find
\begin{equation}\label{brane_stress_one}
T^{\mu}_{\ \nu}[T] = - {\frac{1}{2}}
e^{\Phi}\left({\frac{\Lambda\rho_{\perp}}{\sqrt{-g}}}\right) V(T)
\sqrt{-A}(A^{-1})^{\alpha\mu}g_{\alpha\nu} \, ,
\end{equation}
\begin{equation}\label{hat_tachyon_stress}
{\hat{T}}^{\mu}_{\ \nu} [T] = - {\frac{1}{2}} e^{\Phi}
\left({\frac{\Lambda\rho_{\perp}}{\sqrt{-g}}} \right) V(T)
\sqrt{-A} \left[(A^{-1})^{\alpha\mu}g_{\alpha\nu} -
{\frac{1}{8}}g^{\mu}_{\nu}(A^{-1})^{\lambda\sigma}g_{\lambda\sigma}
\right] \, .
\end{equation}
The other object we need for the metric equation of motion is
\begin{equation}\label{hat_Ramond_stress}
{\hat{T}}^{\mu}_{\ \nu}[C] = {\frac{1}{2(p+1)!}} e^{2\Phi}
G^{\mu\lambda\ldots\sigma} G_{\nu\lambda\ldots\sigma}
-{\frac{(p+1)}{16}} e^{2\Phi} {\frac{G^{2}}{(p+2)!}} g^{\mu}_{\
\nu} \, .
\end{equation}
Lastly, for the tachyon we find the equation of motion
\begin{equation}\label{Tachyon_bulk}
{\frac{dV}{dT}}e^{-\Phi}\sqrt{-A} - \partial_{\mu}
\left[V(T)e^{-\Phi} \sqrt{-A}(A^{-1})^{\mu\alpha}\partial_{\alpha}T
 \right]
+ f(T)\epsilon^{\mu\ldots\lambda}G_{\mu\ldots\lambda} =0 \, .
\end{equation}
In the Discussion section we will make some remarks about the
robustness of these equations.

%--------------------------------------------------------------------+
\subsection{The homogeneous  brane self-consistent
ansatz}\label{subsection:specific_ansatz}

As pointed out earlier, we are interested in time-dependent
processes by which massless Type IIa or Type IIb supergravity
fields are sourced by an open-string tachyon mode on the
worldvolume of an unstable brane. A reasonable assumption is that
the gravitational background generated by backreaction of the
rolling tachyon is of the form
\begin{equation}\label{genmet}
ds^{2} = -dt^{2} + a(t)^{2} d\Sigma^{2}_{p+1}(k_\parallel) +
R(t)^{2} d\Sigma^{2}_{8-p}(k_\perp) \, ,
\end{equation}
where the $n$-dimensional Euclidean metric $d\Sigma^2_{n}(k)$ is
\begin{equation} \label{little}
d\Sigma^2_{n}(k) =\left\{
\begin{array}{ll}
%\vphantom{\sum_{i=1}^{n}}
d\Omega^2_{n}& {\rm for}\; k = +1\\
dE_n^2& {\rm for}\; k = 0 \\
%\vphantom{\sum_{i=1}^{n}}
d H^2_{n} &{\rm for}\; k = -1\ ,
\end{array} \right.
\end{equation}
where $d\Omega^2_{n}$ is the unit metric on $S^{n}$, $dE_n^2$ the flat
Euclidean metric, and $dH^2_{n}$ the `unit metric' on $n$--dimensional
hyperbolic space $H^{n}$.  The corresponding symmetry groups are
\begin{equation}\label{symgps}
\left\{
\begin{array}{ll}
%\vphantom{\sum_{i=1}^{n}}
SO(n+1) & {\rm for}\; k = +1\\
ISO(n) & {\rm for}\; k = 0 \\
%\vphantom{\sum_{i=1}^{n}}
SO(1,n) &{\rm for}\; k = -1\ .
\end{array} \right.
\end{equation}
For $k{=}\pm 1$ we obviously require that $n\ge 2$.

We note that the ansatz (\ref{genmet}), for $k_{\perp}=-1$ and
$k_{\parallel}=0$, appears, after using an appropriate change of
coordinates, to be equivalent to that considered for supergravity
SD-branes in ref.~\cite{KMP}. One can show that this is actually
{\em not} the case. In order to bring solutions of the form
(\ref{genmet}) with $-\infty < t < +\infty$ to the form introduced
in ref.~\cite{KMP} we must find a change of coordinates such that
\beq dt^{2} = d\tau^{2} \,
F(\tau)^{1/2}(\beta(\tau)\alpha(\tau))^{\frac{2}{7-p}}\left(
\frac{\beta(\tau)}{\alpha(\tau)}
\right)^{(k_{1}+\tilde{k})(p-1)/(7-p)}, \eeq where $-\infty < \tau
< +\infty$. It turns out that for all values of the parameters
associated with the isotropic supergravity solutions, such a
change of coordinates does not exist. However, the main feature of
our analysis is that we are allowing for modifications of SD-brane
physics in the region close to the spacelike worldvolume (the
region around $t=0$ for the system of coordinates we use). It
should therefore have been expected that our new solutions are not
included in those presented in ref.~\cite{KMP}.  However, we do
expect the asymptotics to agree.

To be physically relevant, solutions should be asymptotically
flat. For example, SD-brane gravity solutions will be such that
\begin{equation}
\lim_{t\rightarrow \pm\infty} \dot{a}(t) = 0\, , \quad
\lim_{t\rightarrow \pm\infty} \dot{R}(t) = 1\, ,
\end{equation}
for $k_{\parallel}=0$ and $k_{\perp}=-1$. We will also see in our
numerical analysis that only some values of $k_{\perp}$ and
$k_{\parallel}$ are allowed. Also, we expect that for the dilaton
and R-R field
\begin{equation}
\lim_{t\rightarrow\pm\infty} {\dot{C}}(t) =0\,, \quad
\lim_{t\rightarrow\pm\infty} {\dot{\Phi}}(t) =0.
\end{equation}
By inspection of the tachyon equation of motion (\ref{Tachyon_bulk}),
we see that the electric- (or magnetic-) only ansatz referred to at
the beginning of this section will be obviously consistent if we only
allow worldvolume time-derivatives.  This is tantamount to imposing an
$ISO(p+1)$ symmetry on the worldvolume. Ref.~\cite{larsen} argues that
spatial inhomogeneities of the tachyon field will play an important
role in the decay (a view which is also supported, although using a
different line of reasoning, by the results of
ref.~\cite{hashimoto,cosmotachyonK}).  It will be interesting to
investigate the full importance of such effects in the context of our
effective supergravity analysis.  We will include a discussion of the
nontrivial issues raised in the Discussion section.

It turns out that the equations for the combined bulk-brane
evolution in the time-dependent system are complicated enough to
require numerical solution.  For this reason, we will not be able
to accommodate the most natural ansatz\footnote{Strictly speaking,
instead of being a delta-function distribution, the more general
ansatz for the source should be extended ({\it e.g.} a Gaussian)
with its size of the order of the string length.}
$\rho_{\perp}=\delta({\vec{x}})$. Instead, we will use the
``smeared'' ansatz also used by Buchel et al. in
ref.~\cite{buchel},
\begin{equation}\label{Buchel_smear}
\rho_{\perp} =\rho_{0}\sqrt{g_{\perp}} \, .
\end{equation}
Note that in this ansatz, the implicit time dependence in the
transverse metric components is not varied in producing the
equation of motion, rather it is only taken into consideration at
the end of the calculation.  Also, a smeared brane source does not
contribute stress-energy perpendicular to the worldvolume, which
is in the directions $t\,,{\vec{y}}$. It should be noted that the
effect of using this ansatz will be minimized by using a small
value for the density parameter $\rho_{0}$. Of course, the aim
when using such an ansatz is to get rid of any brane action
dependence on the transverse coordinates ${\vec{x}}$.

We should remark that supergravity solutions corresponding to
unstable D-brane systems have been found before \cite{ozetal}.
Their solutions are time-independent, a feature which might seem
rather unreasonable since they are, after all, supposed to
describe unstable objects. Typically, these solutions are nakedly
singular; there is no horizon.  For reasons discussed previously,
these solutions would therefore justifiably be regarded with some
level of suspicion. Possibly, we should really regard these
solutions as fixed-time snap-shots of the unstable brane system
during its evolution.  They do however reflect one desirable
feature: taking into account warping of space in the directions
transverse to the unstable branes.

What we really want is a sort of hybrid of that approach -- where
transverse dependence is the only dependence -- and what we are doing
here -- where time dependence is all there is.  This is something we
postpone to a future investigation; remarks on this will be given in
the Discussion section.

Let us now get back to the simplified ansatz, and just go ahead
and solve it.  We are therefore interested in the precise system
of {\em ordinary} differential equations for our coupled
tachyon-supergravity system. Using the form $C_{12\ldots
p+1}\equiv C(t)$ (which is consistent with our ansatz) the
equation of motion for the R-R field (\ref{bulk_Ramond-Ramond})
becomes
\begin{equation}\label{RR_ansatz}
{\ddot{C}}+ {\dot{C}}\left[(8-p){\frac{{\dot{R}}}{R}}
-(p+1){\frac{{\dot{a}}}{a}}\right] = \lambda a^{p+1}f(T)
{\dot{T}}.
\end{equation}

Now let us find the dilaton equation of motion. A useful identity
is
\begin{equation}\label{R_identity}
{\cal R} = 5(\partial\Phi)^2 - {\frac{9}{2}}\nabla^2\Phi +
{\frac{(3-p)}{8(p+2)!}} e^{2\Phi}G^2 + {\frac{1}{8}}\left(
{\frac{\Lambda\rho_\perp}{\sqrt{-g}}} \right)
e^{\Phi}V(T)\sqrt{-A}(A^{-1})^{\lambda\sigma}g_{\lambda\sigma} \,
,
\end{equation}
with which the dilaton equation of motion can be written,
\begin{equation}
2(\partial\Phi)^2 - \nabla^2\Phi = {\frac{(p-3)}{4(p+2)!}}
e^{2\Phi}G^2 + \left( {\frac{\Lambda\rho_\perp}{\sqrt{-g}}}
\right) e^{\Phi}V(T)\sqrt{-A} \left[ 1-{\frac{1}{4}}
(A^{-1})^{\lambda\sigma}g_{\lambda\sigma} \right] \, .
\end{equation}
This last expression is simply the Einstein frame equation of
motion. So the dilaton in our ansatz satisfies
\begin{eqnarray}\label{final_dilaton}
&& {\ddot{\Phi}} +{\dot{\Phi}}
\left[(8-p){\frac{{\dot{R}}}{R}}+(p+1){\frac{{\dot{a}}}{a}}
\right] -2{\dot{\Phi}}^{2} \nonumber\\ && = {\frac{(3-p)}{4}}
\left({\frac{e^\Phi {\dot{C}}}{a^{(p+1)}}}\right)^2
+{\frac{\lambda}{4}} e^{\Phi} V(T) \left[(3-p)\sqrt{\Delta} -
{\frac{1}{\sqrt{\Delta}}} \right] \, .
\end{eqnarray}
This will be used %in the Einstein equations
whenever double time-derivatives of the dilaton need to be
substituted for.

With $T=T(t)$ we find that the dynamics of the tachyon field is
governed by
\begin{equation}\label{Tachyon_motion}
{\ddot{T}} =  \Delta \left\{ {\dot{\Phi}}{\dot{T}} -
{\dot{T}}\left[ (p+1){\frac{\dot{a}}{a}} \right] -
{\frac{1}{V(T)}}{\frac{dV(T)}{dT}} + {\frac{f(T)}{V(T)}} {\dot{C}}
e^{\Phi} a^{-(p+1)}\sqrt{\Delta} \right\} \, ,
\end{equation}
where $\Delta = 1 - \dot{T}^{2}$. In this paper we will be
assuming that $|f(T)|=V(T)$, a statement which has been shown to
be correct only at past and future asymptopia.  However, we have
also done numerical experiments which show that some breaking of
this relation at intermediate times (near the hilltop) does not
change the important features of our solutions.

We now turn to the equations of motion for the metric components
$a(t)$ and $R(t)$. For the stress-tensors, eliminating second
order derivatives in matter fields, we have
\begin{eqnarray}\label{stressed_out}
% time dirn
{\hat{T}}^{t}_{\ t} &\! = &\!
 4({\dot{\Phi}})^{2} -2{\dot{\Phi}} \left[
(p+1){\frac{{\dot{a}}}{a}} + (8-p){\frac{{\dot{R}}}{R}} \right]
+ {\frac{(5-2p)}{4}}
  \left({\frac{e^{\Phi}{\dot{C}}}{a^{p+1}}}\right)^{2}
\cr && \qquad\qquad
+ {\frac{1}{4}}\lambda e^{\Phi} V(T)
  \left[(7-2p)\sqrt{\Delta}-{\frac{4}{\sqrt{\Delta}}} \right] \, ,
\cr
% parallel dirn
{\hat{T}}^{y}_{\ y} &\! = &\!
 2{\dot{\Phi}} {\frac{{\dot{a}}}{a}}
- {\frac{1}{4}}
  \left({\frac{e^{\Phi}{\dot{C}}}{a^{p+1}}}\right)^{2}
- {\frac{1}{4}}\lambda e^{\Phi} V(T) \sqrt{\Delta}  \, , \cr
% transverse dirn
{\hat{T}}^{x}_{\ x} &\! = &\!  2{\dot{\Phi}} {\frac{{\dot{R}}}{R}}
+ {\frac{1}{4}}
  \left({\frac{e^{\Phi}{\dot{C}}}{a^{p+1}}}\right)^{2}
+ {\frac{1}{4}}\lambda e^{\Phi} V(T) \sqrt{\Delta} \, .
\end{eqnarray}
The components of the Ricci tensor are easily
evaluated:
\begin{eqnarray}\label{Ricci_components}
{\cal R}^{t}_{\ t}&=& (p+1){\frac{\ddot{a}}{a}}
+(8-p){\frac{\ddot{R}}{R}} \, , \\
{\cal R}^{y}_{\ y} &=& {\frac{\ddot{a}}{a}} +
(8-p){\frac{\dot{a}}{a}}{\frac{\dot{R}}{R}} + p\left[
\left({\frac{\dot{a}}{a}}\right)^2
+ {\frac{k_\parallel}{a^2}} \right] \, , \\
{\cal R}^{x}_{\ x}&=&{\frac{\ddot{R}}{R}}  +
(p+1){\frac{\dot{a}}{a}}{\frac{\dot{R}}{R}}  +
(7-p)\left[\left({\frac{\dot{R}}{R}}\right)^{2}
+{\frac{k_{\perp}}{R^{2}}}\right] \, .
\end{eqnarray}
For the (string-frame) ``Einstein'' equations, the time,
longitudinal and transverse components are respectively
\begin{eqnarray} \label{magic} (p+1){\frac{\ddot{a}}{a}}
+(8-p){\frac{\ddot{R}}{R}} &= & + 4({\dot{\Phi}})^{2}
-2{\dot{\Phi}} \left[
  (p+1){\frac{{\dot{a}}}{a}} + (8-p){\frac{{\dot{R}}}{R}} \right]
\cr &&
+ {\frac{(5-2p)}{4}}
  \left({\frac{e^{\Phi}{\dot{C}}}{a^{p+1}}}\right)^{2}
  + {\frac{1}{4}}\lambda e^{\Phi} V(T)
    \left[(7-2p)\sqrt{\Delta}-{\frac{4}{\sqrt{\Delta}}} \right] \,
    , \end{eqnarray}
\beq \label{aeq} {\frac{\ddot{a}}{a}} =  -
(8-p){\frac{\dot{a}}{a}}{\frac{\dot{R}}{R}} -
p\left[\left({\frac{\dot{a}}{a}}\right)^2
       + {\frac{k_\parallel}{a^2}} \right]
+ 2{\dot{\Phi}} {\frac{{\dot{a}}}{a}} - {\frac{1}{4}}
  \left({\frac{e^{\Phi}{\dot{C}}}{a^{p+1}}}\right)^{2}
- {\frac{1}{4}}\lambda e^{\Phi} V(T) \sqrt{\Delta} \, , \eeq
\beq \label{req} {\frac{\ddot{R}}{R}} = -
(p+1){\frac{\dot{R}}{R}}{\frac{\dot{a}}{a}} -
(7-p)\left[\left({\frac{\dot{R}}{R}}\right)^{2}
             +{\frac{k_{\perp}}{R^{2}}}\right]
+ 2{\dot{\Phi}} {\frac{{\dot{R}}}{R}} + {\frac{1}{4}}
  \left({\frac{e^{\Phi}{\dot{C}}}{a^{p+1}}}\right)^{2}
+ {\frac{1}{4}}\lambda e^{\Phi} V(T) \sqrt{\Delta} \, . \eeq
In the end we have a system of five second order coupled ordinary
differential equations for $\{T,C,\Phi,a,R\}$ as functions of $t$.
These are respectively equations (\ref{Tachyon_motion}),
(\ref{RR_ansatz}), (\ref{final_dilaton}), (\ref{aeq}) and
(\ref{req}). For consistency this system of equations must be
supplemented with the first-order constraint
\beq\begin{array}{l} \label{fconstraint}
{\displaystyle{
\frac{1}{\lambda e^{\Phi}}\left\{
2(p+1)(8-p)\frac{\dot{a}}{a}\frac{\dot{R}}{R} + p(p+1) \left[
\left(\frac{\dot{a}}{a}\right)^{2}+\frac{k_{\parallel}}{a^{2}}\right]
+(7-p)(8-p)\left[
\left(\frac{\dot{R}}{R}\right)^{2}+\frac{k_{\perp}}{R^{2}}\right]
\right. }}
 \nonumber \\ \qquad\qquad {\displaystyle{ \left. -4\dot{\phi}\left[
(p+1)\frac{\dot{a}}{a}+(8-p)\frac{\dot{R}}{R}-\dot{\phi}\right]-\frac{1}{2}
\left(\frac{e^{\Phi}\dot{C}}{a^{p+1}}\right)^{2}\right\} }}
={\displaystyle{
\frac{V(T)}{\sqrt{\Delta}} }}
\end{array}\eeq
obtained by plugging eqs.~(\ref{aeq}) and (\ref{req}) in
eq.~(\ref{magic}). Of course, if this last equation is satisfied at
$t=0$ it will be for all times.

%====================================================================+
\section{The roll of the tachyon}
\label{section:numeric}

We refer to a supergravity SD-brane as the field configuration
generated by a system composed of a large number, $N$, of
microscopic SD-branes. As emphasized in section
\ref{subsection:specific_ansatz}, there is a single continuous
parameter that controls the dynamics of these fields, {\it i.e.},
$\lambda = \rho_{0} \Lambda$. This is the parameter that
determines the relative importance of the unstable brane source.
Clearly, for $\lambda\rightarrow 0$ the open string tachyon
decouples from the bulk fields (no backreaction) and the
corresponding supergravity solutions will presumably be the
singular ones found in refs.~\cite{gutperle,gutperle2,KMP}.

In this section we present solutions with the symmetries of
SD-branes and a non-vanishing $\lambda$. We demonstrate that they
are generically non-singular.

%--------------------------------------------------------------------+
\subsection{Tachyon evolution in flat space}

The solutions relevant for SD-brane physics in Type II a,b
superstring theory could be the ones corresponding to an open
string tachyon evolving from \beq \label{bc5} \lim_{t\rightarrow
-\infty} T(t) = +\infty \eeq to \beq \lim_{t\rightarrow +\infty}
T(t) = -\infty \eeq in a symmetric runaway potential of the form
shown on figure \ref{potential}. Solutions of this type correspond
to the tachyon evolving between two different minima of the
potential. Possible initial conditions (at $t=0$) for these
solutions are of the form \beq \label{bc7} \dot{T}(0) = {\rm
const.} \, , \;\;\;\;\;\; T(0)=0 \, . \eeq Another set of
solutions corresponds to a tachyon evolution with \beq \label{frg}
\lim_{t\rightarrow \pm\infty} T(t) = \alpha \, \infty \, , \eeq
where $\alpha=1$ and $\alpha=-1$ lead to equivalent solutions.
Appropriate boundary conditions for the tachyon (at $t=0$) are
then of the form \beq \label{bc8} \dot{T}(0) = 0 \, , \;\;\;\;\;\;
T(0)= {\rm const.} \eeq In this section we characterize the
supergravity solutions generated by a tachyonic source with the
properties mentioned above. The results we present are for
solutions with boundary conditions (\ref{bc8}) and asymptotic
behavior (\ref{frg}). We have found that the solutions associated
with the boundary conditions (\ref{bc7}) and asymptotic behavior
(\ref{bc5}) have similar qualitative features. Strictly speaking,
our approach to studying real-time evolution in supergravity can
also be extended to more general cases, {\it i.e.}, brane decay or
creation (half SD-branes). \FIGURE{\epsfig{file=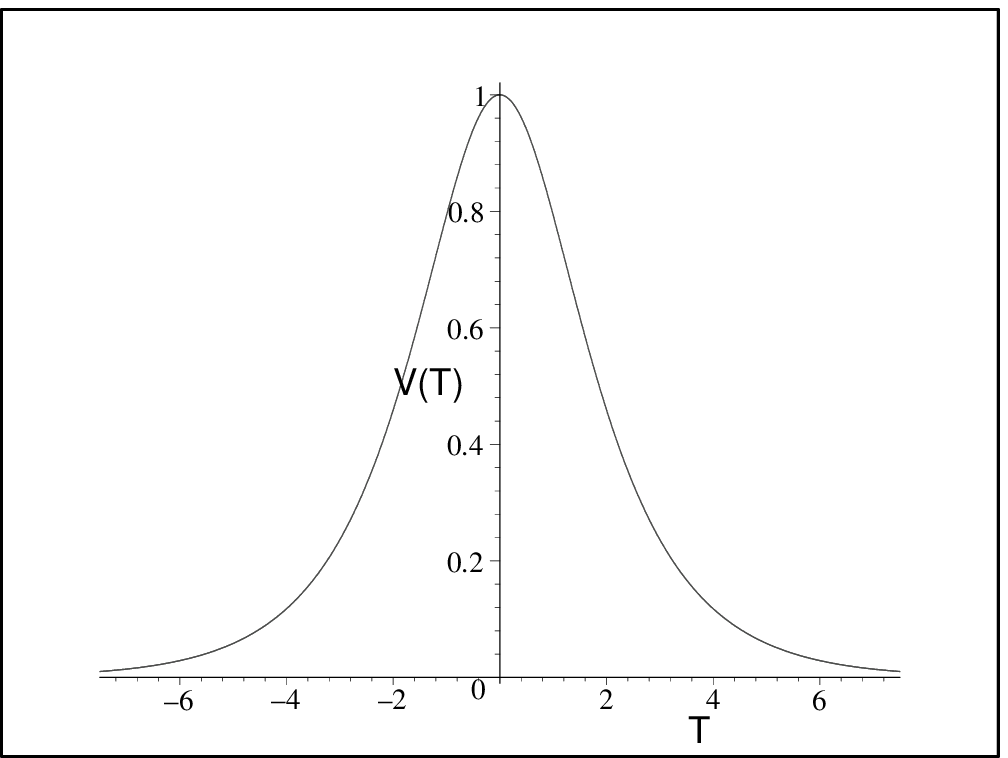,
width = 10cm}\caption{Possible form of the open string tachyon
potential $V(T)$ on a SD-brane in Type II a,b superstring theory.}
\label{potential}}

In our analysis we use the potential \beq \label{pot12}
V(T)=\frac{1}{\cosh \left(T/\sqrt{2}\right)}\eeq because it agrees
with open string field theory calculations for large values of the
tachyon for unstable D-brane systems in Type II a,b superstring
theory.\footnote{In bosonic string theory the potential is
asymmetric and unbounded from below as $T\rightarrow -\infty$.} It
is not known what the exact potential is for intermediate times
but our solutions are only mildly affected by its particular form.
The expression for the tachyon when $\lambda=0$, {\it i.e.}, when
there are no couplings to the bulk supergravity modes,
is\footnote{We refer the reader to Appendix \ref{flatach} for a
derivation of this expression.}
\beq \label{flatfield} T(t) = -\sqrt{2}\;{\rm arcsinh} \left(
-\dot{T}(0) \sinh \frac{t}{\sqrt{2}} \right) \eeq
for the boundary conditions (\ref{bc7}). When the boundary
conditions are of the form (\ref{bc8}), the analytic expression
for the tachyon is \beq \label{flatfield2} T(t) = -\sqrt{2}\;{\rm
arcsinh} \left( \sinh \left( -\frac{T(0)}{\sqrt{2}} \right) \cosh
\frac{t}{\sqrt{2}} \right). \eeq Figure \ref{T} shows a tachyon
profile for $T(0)=0.83$ and $\dot{T}(0)=0$. Homogeneous solutions
such as eqs.~(\ref{flatfield}) and (\ref{flatfield2}), derived
from a tachyonic action, correspond to a fluid that has a constant
energy density and vanishing pressure asymptotically (tachyon
matter). We will shortly see how these features are affected by
the inclusion of couplings to bulk modes.
\FIGURE{\epsfig{file=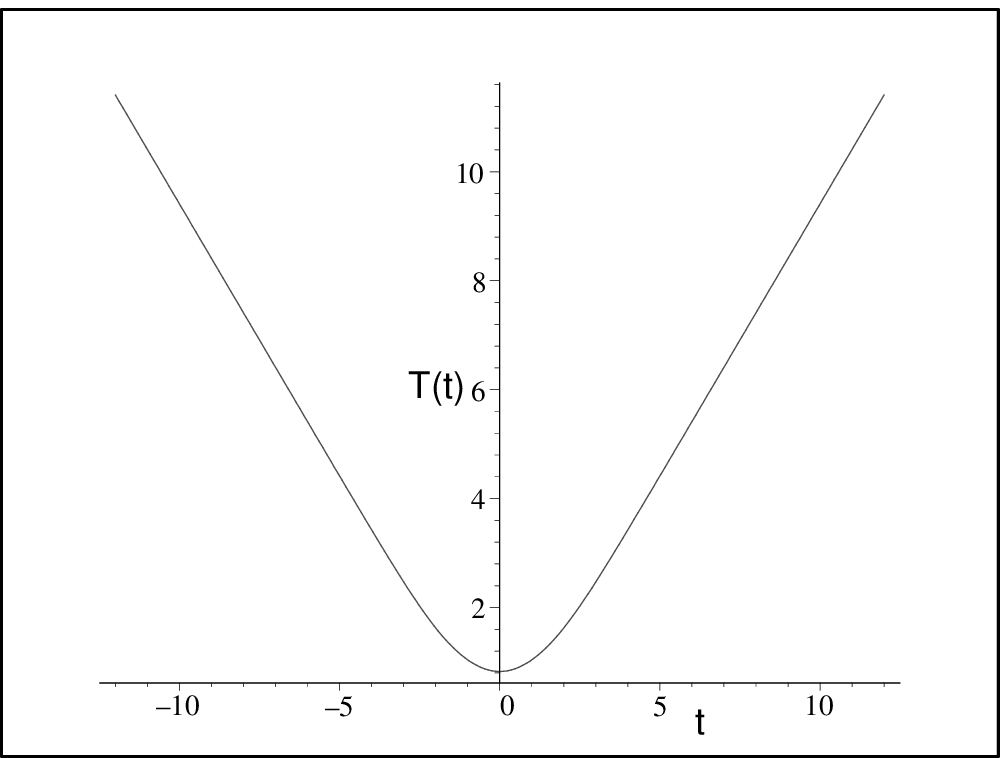, width = 10cm}\caption{The tachyon
field evolution for $\dot{T}(0)=0$ and $T(0)=0.83$.} \label{T}}

\subsection{R-symmetry group}
Before presenting the numerical results we comment on the issue of
R-symmetry. As pointed out in ref.~\cite{gutperle}, SD-brane
solutions should be invariant under the transverse Lorentz
transformations leaving the location of the brane fixed. This
corresponds to an $SO(1,8-p)$ R-symmetry.\footnote{The
interpretation in terms of R-symmetry is relevant for the idea
that SD-brane gravity fields are holographically dual to a
worldvolume Euclidean field theory.} This property is embodied in
the supergravity solutions found in
refs.~\cite{gutperle,gutperle2,KMP} where the transverse space
metric has a factor of the form: $R(t)^{2} dH_{8-p}^{2}$. The
embedding group of the hyperbolic space $H^{8-p}$ is
$SO(1,8-p)$.\footnote{The intuition for the nature of the
R-symmetry group is inherited from the AdS/CFT correspondence. For
example, the metric of a 3-brane is of the form: $ds^{2}=
f(r)\left[ -dt^{2} + d{\vec y}^{2} \right] + 1/f(r)\left[ dr^{2} +
r^{2} d\Omega_{5}^{2} \right]$. The R-symmetry group in this case
is $SO(6)$, a statement that can be traced to the fact that the
near horizon limit of the geometry is AdS$_{5}\times S^{5}$, {\it
i.e.}, the gravitational background dual to the worldvolume field
theory on an ensemble of D3-branes.} This explains why we study
supergravity solutions with $k_{\parallel}=0$ and $k_{\perp}=-1$
in more details.
However, the cases with $k_{\perp}=0$ are also candidate solutions
for SD-brane (and, more generally, unstable D-brane) supergravity
solutions. We present an analysis of those and other cases in
section \ref{othercases}.

It is suggested in ref.~\cite{gutperle} that the spacelike naked
singularities \cite{gutperle,gutperle2,KMP} associated with the
supergravity SD-branes could be resolved by using a metric ansatz
that allows for a breaking of the R-symmetry in the region around
the core of the object ($t=0$). The intuition from the AdS/CFT
correspondence comes from ref.~\cite{nunez} which describes cases
of spontaneously broken R-symmetry. Our ansatz as it is cannot
accommodate such an R-symmetry breaking. Presumably, this would
correspond to a time-dependent process by which the R-symmetry is
broken down to $SO(8-p)$ in a finite region. The closest our
ansatz can come to realizing this scenario is if we consider the
cases with $k_{\perp}=0$. Then, the transverse symmetry group is
$ISO(8-p)$, the compactification of which is $SO(8-p)$. We find
that solutions with $k_{\perp}=0$ are {\it regular} and
asymptotically flat. Moreover, as we will see in
section~\ref{othercases}, a generic feature of the $k_{\perp}=0$
solutions is that the effective string coupling
$g_{s}\exp(\Phi(t))$ vanishes asymptotically. We will also remark
on the cases with $k_\perp$ and/or $k_\parallel$ equal to +1,
which all develop big-crunch singularities in finite time.
%

%--------------------------------------------------------------------+
\subsection{Numerical SD-brane solutions}

Our ansatz for describing supergravity SD$p$-brane solutions is
\begin{eqnarray} ds^{2}_{Sp} &= & -dt^{2} + a(t)^{2}d{\vec y}^{2}
+ R(t)^{2} dH_{8-p}^{2}, \nonumber \\
C & =& C(t)\,,\quad \Phi = \Phi(t),
\end{eqnarray}
where we have taken $k_{\parallel}=0$ and $k_\perp=-1$ in accord
with the original paper ref.~\cite{gutperle}.  Again, these
supergravity bulk modes will be excited by couplings to an
homogeneous open string tachyon as described in section
\ref{section:prelim}. We consider only homogeneous solutions ({\it
i.e.}, depending only on time).  As discussed earlier, it is
possible that inhomogeneities might play an important role in the
creation/decay process \cite{larsen}, but we are postponing this
issue for now.

The resulting system of coupled differential equations for which
we will find solutions is of course highly non-linear. Among other
things, this implies that it is not possible to extract scaling
behavior for the fields from their equations of motion. We could
therefore expect that the behavior of these solutions depends in a
physically important way on the initial conditions for the field
components involved: $a(t)$, $R(t)$, $C(t)$, $\Phi(t)$ and the
source $T(t)$.
We will see that the qualitative behavior of the supergravity
SD-brane solutions actually does not depend very much on these
initial conditions.

The multiplicity of potential solutions could be large: one
solution would be expected for each set of initial conditions.
Now, for SD-branes the number of solutions is greatly restricted
by the constraint equation (\ref{fconstraint}).  It would appear
natural to consider, as candidate solutions for SD-branes, those
which are time-reversal symmetric around $t=0$.\footnote{We thank
Alex Buchel for pointing out that the time-reversal symmetric
solutions are inconsistent.  In fact, the constraint equation
(\ref{fconstraint}) cannot be satisfied (for the case
$k_{\parallel}=0$ and $k_{\perp}=-1$) with the boundary
conditions: $\dot{a}(0)=0$, $\dot{R}(0)=0$, $\dot{\phi}(0)=0$,
$\dot{C}(0)=0$ and $T(0)=0$.}  In fact, precisely time-symmetric
solutions actually do not exist! That is, unless they have
$k_\parallel=0,k_\perp=+1$, in which case they have R-symmetry
which is markedly in conflict with the proposals of earlier
papers, e.g. \cite{gutperle}.  As we said above, these solutions
with wrong R-symmetry also develop a big-crunch singularity in
finite time.

In a sense, motivated by inflationary cosmology, the absence of
precisely time-symmetric solutions is not particularly bothersome
to us.  What we mean by this is that it is an extremely fine-tuned
situation to have exactly zero kinetic energy in each of the bulk
fields at $t=0$.  Any small quantum fluctuation of a bulk field
takes us away from time-symmetry.  Therefore, all solutions which
we will exhibit here will have some kinetic energy in one or more
of the bulk fields at $t=0$. And the kick in bulk field(s) at
$t=0$, required to solve the initial constraint, can actually be
made very small by, {\it e.g.}, choosing $R(0)$ large. Such
initial conditions are quite generic: even a little bit of kick in
$a(t)$ alone will suffice to give completely nonsingular solutions
for the entire time-evolution.  The precisely time-symmetric
solutions are however inaccessible; see subsection \ref{trssblar}
for a more detailed exposition of this case.

Another reason why we find a slight bulk asymmetry at $t=0$ reasonable
relates to particle/string production.  We have of course neglected
such production in our analysis.  It is nonetheless clear that, for a
full SD-brane solution, backreaction will combine with particle
production to make bulk fields naturally {\em a}symmetric.  It is only
in the approximation of zero backreaction that time-symmetry is
possible when particle production occurs, but of course that
approximation cannot be self-consistent.

Let us now move to the solution of our problem of interest.
Unfortunately we have been unsuccessful at finding analytical
expressions for the bulk modes and tachyon field when their mutual
coupling is non-vanishing ($\lambda\neq 0$). We therefore resorted to
solving the corresponding system of differential equations
numerically.\footnote{Recall that this is the primary reason why it is
so difficult to include ${\vec{x}}$-dependence in our ansatz.}
In what follows we show the results associated with a SD4-brane,
and present some interesting analysis of the effect of varying the
initial conditions. We provide general comments for other
SD-branes with $p<7$ and explain how the SD$p$-branes with $p=7,8$
are different. We comment on the robustness of the solutions to
changes in initial conditions. Throughout our analysis we pay
special attention to the impact of varying the initial condition
$\Phi(0)$ on the dilaton, because this controls the string
coupling close to the hilltop.

For finding the numerical solutions we fix the initial conditions
at $t=0$, and evolve this data forward in time. Then, we evolve
the same data backward in time from $t=0$. The result is the
solution associated with a full SD-brane, {\it i.e.}, the bulk
fields sourced by the open string tachyon as it evolves from
$t=-\infty$ to $t=+\infty$. In this section we present the results
associated with a SD4-brane with boundary conditions
\begin{eqnarray} \label{fullbc} T(0)=0.83 \, , \;\;\;
\dot{T}(0)=0 \, , \;\;\; a(0)=1
\, , \;\;\; \dot{a}(0)= 0.091 \, , \;\;\; R(0)=10 \, , \nonumber
\\ \dot{R}(0) = 0.01 \, , \;\;\; \phi(0)=-1 \, , \;\;\;
\dot{\phi}(0) =0.01 \, , \;\;\; C(0) = 0 \, , \;\;\; \dot{C}(0) =
0.01 \, , \end{eqnarray}
and $\lambda = 0.1\, $. Again, these solutions correspond to the
tachyon rolling up the potential from $T=+\infty$ ($t=-\infty$)
and coming to a halt for $T(0)=0.83$ which corresponds to a
turning point. Then, the tachyon evolves toward the bottom of the
potential at $T=+\infty$ for $t=+\infty$.  This type of tachyon
evolution was considered, for example, in
refs.~\cite{maldacena,andylastweek} where they are called full
S-branes. We could of course have chosen different initial
conditions - including some with less time-asymmetry; these are
used just for the sake of illustration.
%

%- - - - - - - - - - - - - - - - - - - - - - - - - - - - - - - - - - +
\subsubsection{Deformation of the tachyon field}

Figure~\ref{T} shows the evolution of the tachyon for the
S4-brane. For $\lambda=0$ (no coupling between the closed and the
open string modes), we find \beq \lim_{t\rightarrow \pm \infty}
T(t) = \pm t. \eeq This observation is suggestive that the tachyon
field might play the role of time itself in cosmological models
driven by brane decay (as proposed in ref.~\cite{sen1}). We find
that this asymptotic behavior for the tachyon survives when
couplings to bulk modes are introduced. In fact, for $\lambda\neq
0$ we find \beq \lim_{t\rightarrow \pm\infty} T(t) = \pm (t +
\kappa_{\scriptscriptstyle \pm T}). \eeq The constants
$\kappa_{\scriptscriptstyle \pm T}$ depend non-trivially on the
boundary conditions at $t=0$ and $p$. For example, as $p$ is
increased $\kappa_{\scriptscriptstyle T}$ decreases. Also, for
large values of $\Phi(0)$ the tachyon deformation from the flat
space case becomes larger. As mentioned in Appendix~\ref{flatach},
for $\lambda=0$ the state of the tachyon for large time
($t\rightarrow \pm \infty$) is that of a perfect fluid with
constant energy density and vanishing pressure. This is the
so-called tachyon matter. We find that for $\lambda\neq 0$, both
the energy density and the pressure (physical quantities measured
in the Einstein frame) vanish. In other words, the tachyon matter
is clearly only an illusion of the $g_{s}\rightarrow 0$ limit.

We consider briefly the effect of varying the initial conditions
$\dot{T}(0)$ and $T(0)$. For half SD-branes ({\it i.e.}, the
future of the full SD-branes) we find that the time it takes the
tachyon to reach the bottom of its potential increases for smaller
values of $\dot{T}(0)$ and $T(0)$. Not only that, for very small
initial velocities the tachyon stays perched at the top of the
potential for a certain period of time. In general, we observe
that it takes less time for the tachyon to reach the bottom of its
potential when we increase $\lambda$. Now, we also observe that
the difference between the curves associated with flat space
($\lambda =0$) and $\lambda\neq 0$ tachyons decreases as
$\dot{T}(0)$ and $T(0)$ increase. Also, for large negative values
of $\Phi(0)$, $\kappa_{\scriptscriptstyle T}$ becomes very small.
This is simply a reflection of the fact that such cases correspond
to a very small initial string coupling (see below for details).

There is another interesting feature of the tachyon when coupled
to the massless closed string modes. Firstly, the time it takes
the tachyon to reach the bottom of its potential is not
significantly altered even when considering large values of the
`coupling' $\lambda$. Finally, we find that as the coupling
$\lambda$ is increased, $\kappa_{\scriptscriptstyle T}$, or the
deformation away from the flat space tachyon, increases
correspondingly.

%- - - - - - - - - - - - - - - - - - - - - - - - - - - - - - - - - - +
\subsubsection{Time-dependent string coupling}\label{subsec:dildil}

The string coupling is given by the expression \beq g_{s} =
e^{<\Phi_{0}>} \,, \eeq where $\Phi_{0}$ is the background dilaton
field in the absence of sources, {\it i.e.}, strings and D-branes.
Typically the presence of stringy excitations will modify the
coupling of the theory, \beq g_{s} \rightarrow g_{s}
e^{\Phi(\eta)}, \eeq where $\eta$ is a spacelike variable for
D-branes and is timelike for SD-branes.

For supergravity D$p$-brane solutions the dilaton field is (see,
for example, ref.~\cite{peet}), \beq g_{s}(r) = g_{s} e^{\Phi(r)}
= g_{s} \left( 1+ \frac{c_{p}g_{s}N_{p}l_{s}^{7-p}}{r^{7-p}}
\right)^{\frac{1}{4}(3-p)}, \eeq where
$c_{p}=(2\sqrt{\pi})^{5-p}\Gamma[\frac{1}{2}(7-p)]$. The string
coupling is seen to vary according to whether test closed strings
propagate close or far from the horizon ($r=0$) of these
geometries. For all static supergravity solutions (including
NS5-branes), the asymptotic string coupling is such that\beq
\lim_{r\rightarrow +\infty} g_{s}(r) = g_{s}. \eeq The effect of
supergravity D-branes is therefore to modify the coupling locally.
For $p<3$, it is large close to the horizon and decreases to
$g_{s}$ as $r\rightarrow +\infty$. For $p>3$, the string coupling
is small close to the horizon but increases to $g_{s}$ for large
$r$. The case $p=3$ is special because the dilaton field sourced
by the 3-brane is constant throughout the spacetime. Typically,
the size of the region where the dilaton is not constant depends
on the parameter $g_{s}N$. Large values of this parameter are
associated with larger regions where dilaton perturbations
associated with the brane are noticeable.

The solutions associated with supergravity SD-branes induce
dilaton perturbations corresponding to a time-dependent string
coupling, \beq g_{s}(t) = g_{s} e^{\Phi(t)}. \eeq We find that the
time dependence of the dilaton sourced by SD-branes is
qualitatively different when compared to the radial dependence of
the dilaton associated with regular D-branes.\footnote{This is an
other example where features of SD-branes are not simply those
inherited by analytic continuation of D-branes. In fact, a double
analytic continuation ($r\rightarrow i r$ and $t\rightarrow i t$)
of the supergravity D$p$-brane solutions lead to objects with an
imaginary R-R charge. This is unphysical.}

Figure~\ref{dilaton} shows the time-evolution of the dilaton
function $g_{s}^{-1} e^{\Phi(t)}$ for a SD4-brane with boundary
conditions (\ref{fullbc}). Typically, the function
$g_{s}^{-1}e^\Phi(t)$ decreases from $t=0$ as $t\rightarrow \pm
\infty$. We find that smaller values of $p$ correspond to larger
asymptotic string couplings.

The dilaton field generated by SD-branes has at least two
interesting properties. First, all solutions are such that the
dilaton stabilizes to a constant asymptotically, \beq
\lim_{t\rightarrow \pm \infty} \Phi(t) = \Phi_{\pm\infty}. \eeq
More generally, the relation between $\Phi_{+\infty}$ and
$\Phi_{-\infty}$ depends on the initial conditions. This last
statement applies to all other bulk fields. Secondly, the
asymptotic value of the dilaton is always smaller than its initial
value at $t=0$, \beq \Phi_{\pm\infty} < \Phi(0). \eeq This implies
that the late/early string coupling ($t\rightarrow \pm \infty$) is
always smaller than the coupling when the tachyon is at the top of
its potential ($t=0$), {\it i.e.}, \beq g_{s} e^{\Phi_{\pm\infty}}
< g_{s} e^{\Phi(0)}. \eeq

We find that the qualitative features shown on figure
\ref{dilaton} are preserved when the boundary conditions on the
various fields are changed. Nevertheless, we consider the effect
of varying $\Phi(0)$ in some detail. An interesting quantity to
study is the ratio \beq h =
\frac{e^{\Phi_{\pm\infty}}}{e^{\Phi(0)}}, \eeq which gives a
quantitative measure of how much the initial string coupling is
modified asymptotically. The tachyon profile is not altered
significantly when the initial condition on $\Phi(t)$ is varied.
Nevertheless, we observe that that for large values of $\Phi(0)$
(large initial coupling) the tachyon field reaches the bottom of
the potential well faster. A large initial coupling also means
that the bulk fields relax faster to their stable asymptotic
configuration compared to cases where $\Phi(0)$ is
smaller.\footnote{Obviously $\Phi(0)$ can be taken to have
negative values.} The overall effect on the bulk fields is also
enhanced for larger values of $\Phi(0)$. For example, as the
initial coupling is increased the scale factor $a(t)$ stabilizes
to significantly smaller values (see next section). As for the
dilaton field itself, we find that the ratio $h$ is large for
larger values of $\Phi(0)$. For very small values of the initial
string coupling (large negative values of $\Phi(0)$), the ratio
$h$ approaches unity.

In summary, we find that the parameter $\Phi(0)$, {\it i.e.}, the
parameter determining the string coupling when the tachyon is
close to the top of its potential, strongly determines the {\it
importance} of the unstable brane source effect on the
supergravity bulk modes.

\FIGURE{\epsfig{file=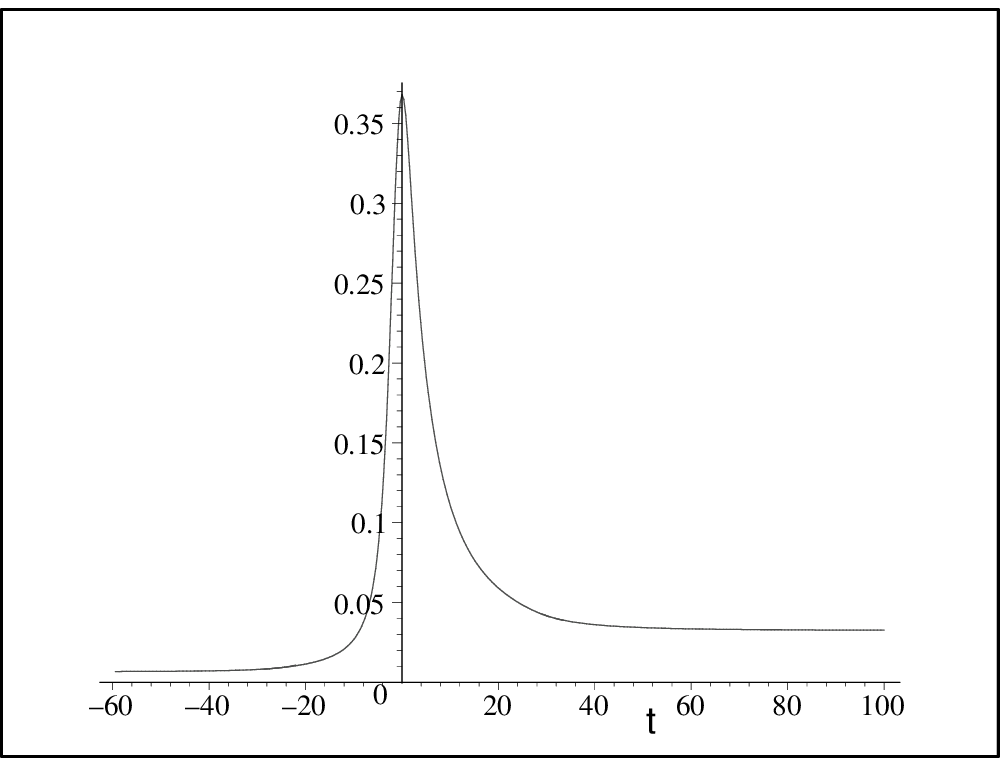}\caption{Time dependence of the
function $e^{\Phi(t)}/g_{s}$ for a SD4-brane with boundary
conditions (\ref{fullbc}).} \label{dilaton}}

%- - - - - - - - - - - - - - - - - - - - - - - - - - - - - - - - - - +
\subsubsection{Gravitational field}

We now describe the effect of the unstable brane source on the
time-dependent metric components $a(t)$ and $R(t)$.

\FIGURE{\epsfig{file=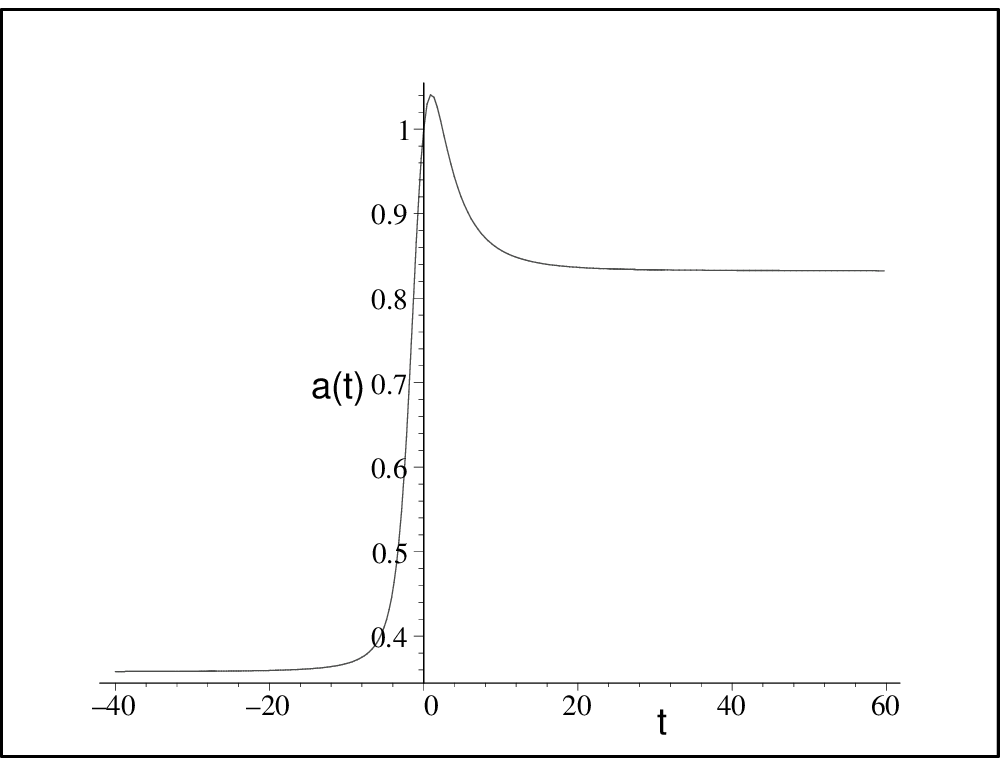, width = 10cm}\caption{The scale
factor on the worldvolume of a supergravity SD4-brane with
boundary conditions (\ref{fullbc}).} \label{a}}

Figure \ref{a} shows the scale factor $a(t)$ on the worldvolume of
a SD4-brane with the boundary conditions (\ref{fullbc}). A general
feature is that away from $t=0$ the scale factor monotonically
decreases from its initial value, $a(0)$, to a stable asymptotic
value, \beq \lim_{t\rightarrow \pm \infty} a(t) = a_{\pm\infty},
\eeq with $a_{+\infty}>a_{-\infty}$. The time it takes for  this
scale factor to reach its asymptotic value corresponds roughly to
the time it takes for the tachyon to reach the bottom of its
potential. As pointed out above, for large initial values of the
dilaton, $\Phi(0)$, the asymptotic values of the scale factor,
$a_{\pm\infty}$, become smaller. Correspondingly, when the initial
coupling is weak the effect of the probe on the bulk modes is
small and $a(t)$ stabilizes to a value closer to its initial value
$a(0)$.

Figure \ref{Rt} shows the behavior of the metric function $R(t)$.
A generic feature of the supergravity SD-brane solutions is that
\beq \lim_{t \rightarrow \pm \infty} R(t) = \pm (t +
\kappa_{\scriptscriptstyle \pm R}). \eeq The constants
$\kappa_{\scriptscriptstyle \pm R}$ are generically larger for
larger values of $p$.

\FIGURE{\epsfig{file=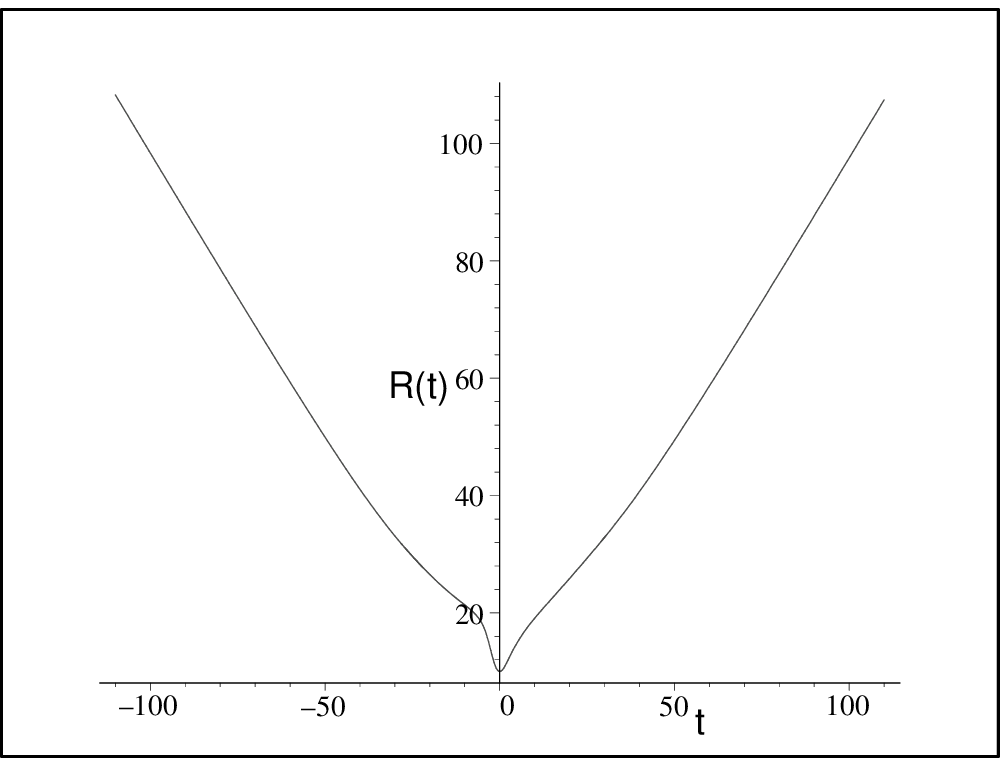}\caption{The SD4-brane
transverse scale factor $R(t)$ with boundary conditions
(\ref{fullbc}).} \label{Rt}}

%- - - - - - - - - - - - - - - - - - - - - - - - - - - - - - - - - - +
\subsubsection{Ramond-Ramond field}

Figure~\ref{Ct} shows the time dependence of the Ramond-Ramond
form field for a supergravity SD4-brane with boundary conditions
(\ref{fullbc}). Again, the energy stored in this field
(proportional to its time-derivative) goes to zero in
approximately the time it takes for the tachyon to reach the
bottom of its potential. A generic feature of the Ramond-Ramond
field associated with a SD-brane is therefore, \beq
\lim_{t\rightarrow \pm \infty} C(t) = C_{\pm\infty}, \eeq where
$C_{\pm\infty}$ is a constant. Typically we find that these
constants are smaller for larger values of $p$.

\FIGURE{\epsfig{file=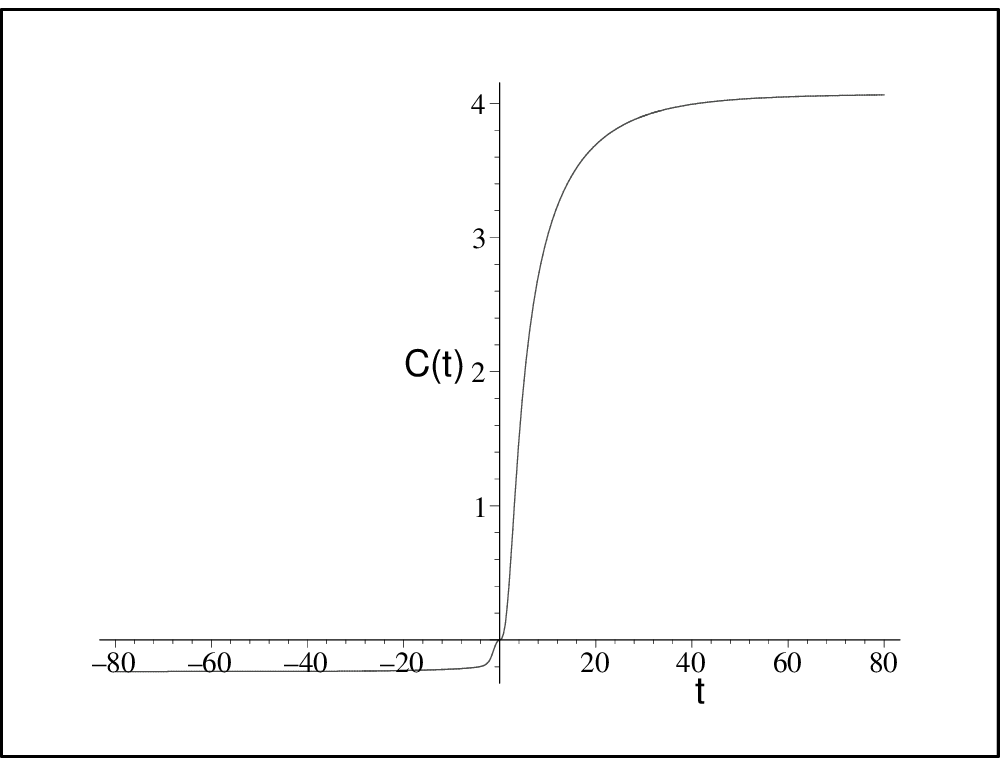}\caption{The SD4-brane
Ramond-Ramond field $C(t)$ associated with the boundary conditions
(\ref{fullbc}).} \label{Ct}}

%- - - - - - - - - - - - - - - - - - - - - - - - - - - - - - - - - - +
\subsubsection{Curvature bounds and asymptotic flatness}

The supergravity equations of motion are derived from a worldsheet
calculation by requiring that, at a certain order in perturbation
theory, the beta-functions associated with bulk fields vanish.
Typically, there are higher order (in $\alpha'$) curvature
corrections to the beta-functions. These corrections are
negligible only if the curvature involved is small when measured
with respect to the string length, $l_{s} = \sqrt{\alpha'}$. This
is why our solutions can, strictly speaking, be trusted only if
the curvature involved is such that
$\left| {\cal R} \right| , \; \left| {\cal R}_{\mu\nu}{\cal
R}^{\mu\nu} \right| , \; \left| {\cal R}_{\mu\nu\rho\lambda}{\cal
R}^{\mu\nu\rho\lambda} \right|$ are small.
We verify that this condition is satisfied by studying the
behavior of the time-dependent Ricci scalar of the supergravity
SD-branes,
\begin{eqnarray}
\label{curvat} {\cal R}(t) = && 2(p+1)\frac{\ddot{a}}{a} +
2(8-p)\frac{\ddot{R}}{R}
+2(p+1)(8-p)\frac{\dot{a}}{a}\frac{\dot{R}}{R} \nonumber\\
&& +p(p+1)
\left(\frac{\dot{a}^{2}}{a^{2}}+{\frac{k_\parallel}{a^2}}\right)
+(8-p)(7-p) \left(
\frac{\dot{R}^{2}}{R^{2}} + \frac{k_{\perp}}{R^{2}} \right),
\end{eqnarray}
where $k_{\perp}=-1$ and $k_\parallel=0$ for the cases of interest
here.

A property of the solutions, which is apparent from studying the
evolution of bulk modes, is that of asymptotic flatness. In fact,
we find \begin{eqnarray} \lim_{t\rightarrow \pm \infty}
ds^{2}_{Sp} = -dt^{2} +
a_{\pm\infty}^{2} d{\vec y}^{2} + (t+\kappa_{\pm R})^{2} dH_{8-p}^{2} \, , \\
\lim_{t\rightarrow \pm \infty} \Phi(t) = \Phi_{\pm\infty},
\;\;\;\; \lim_{t\rightarrow \pm \infty} C(t) = C_{\pm\infty},
\end{eqnarray} where $a_{\pm\infty}$, $\kappa_{\pm R}$, $\Phi_{\pm\infty}$
and $C_{\pm\infty}$ are constants. Both the first- and
second-derivative of the bulk modes vanish asymptotically. The
resulting brane configuration is then clearly flat for
$t\rightarrow \pm \infty$, as it should be.

An important question to answer at this stage is: {\it Which
quantity in the problem sets an upper bound on the curvature for
SD-branes ?} First, we have found that whatever the curvature is
at $t=0$, its absolute value will never exceed it significantly in
the course of the evolution. This is true only for half S-branes
with boundary conditions corresponding to positive derivatives of
the bulk fields. More generally for full S-branes the acceptable
solutions are those with a combination of the boundary conditions
such that \beq \lim_{t\rightarrow 0} \left| {\cal R}(t) \right| \
{\rm{small}}\ . \eeq These requirements are certainly attainable
within the self-consistent supergravity approximation we are
considering --- along with the specific ansatz we introduced in
order to be able to solve the equations numerically.

%- - - - - - - - - - - - - - - - - - - - - - - - - - - - - - - - - - +
\subsubsection{The $p=7$ and space-filling SD-branes}

The case $p=8$ should be special because there is no transverse
space into which ``energy'' can be dissipated. The only bulk
fields involved are then the metric component, $a(t)$, the
dilaton, $\Phi(t)$, and the Ramond-Ramond field, $C(t)$. We find
that the time-derivative of the scale factor, $\dot{a}(t)$,
decreases to zero, as $t\rightarrow \pm \infty$, many orders of
magnitude slower than for $p<7$. The scale factor $a(t)$ also goes
to zero after an infinite time (see section \ref{einstein} for a
physical interpretation) which is to be contrasted with the cases
$p<7$ where $a_{\pm\infty} \neq 0$. One would expect that to be
the source of a curvature singularity at $t=\pm\infty$ but it is
not the case. The Ricci scalar for the space filling SD-brane is
\beq {\cal R}(t) = -72 \left(\frac{\dot{a}}{a}\right)^{2}
-\frac{9}{2} \left(\frac{e^{\Phi}\dot{C}}{a^{9}}\right)^{2} +36
\dot{\Phi} \frac{\dot{a}}{a} + \frac{9}{2}\lambda
e^{\Phi}\Delta^{1/2} V(T). \eeq Curvature singularities are
avoided because all time-derivatives in the problem go to zero
faster than the scale factor as $t\rightarrow \pm \infty$. In
particular, the quantity $e^{\Phi}\dot{C}$ goes to zero faster
than $a(t)$ for large time. Also the string coupling slowly goes
to zero asymptotically ($\Phi_{\pm\infty}\rightarrow -\infty$)!
For the Ramond-Ramond field we find the same behavior as for the
cases $p<7$. The only difference is that the relaxation time of
the bulk modes is many orders of magnitude larger than in the
other cases.

The functions $a(t)$, $\Phi(t)$ and $C(t)$ associated with the
SD7-brane behave in the same way as the space-filling SD-brane. Of
course in this case there is a transverse space and we find that
the relaxation of $\left| R(t)\right|$ to its asymptotic form
$t+\kappa_{\scriptscriptstyle \pm R}$ takes an infinite amount of
time. In other words, it relaxes to its asymptotic form much
slower than for the cases $p<7$.

%--------------------------------------------------------------------+
\subsubsection{Einstein frame} \label{einstein}

Let us end this section with a remark about the physics of our
SD-branes in Einstein frame.  This would be the more natural and more
physical frame to use in discussions of potential uses of rolling
tachyons in the context of cosmology.

The transformation from string frame to Einstein frame involves a
multiplicative factor of $e^{-\Phi/2}$ for $d=10$, which is the
dimension in which we are working here.  Now, we have already
commented at length on the behavior of the time-dependent dilaton
field in section \ref{subsec:dildil}.  The essential physics there
was that the dilaton is biggest at the top of the potential hill;
in particular, it stabilizes in the infinite past and future at a
{constant} value smaller than the initial condition at the
hilltop. Dilaton derivatives also remain small at all times during
the evolution.  Therefore, most of the qualitative features of our
solutions will be preserved upon transformation to Einstein frame;
in particular, all solutions remain completely nonsingular.

A case that deserves further comments is that of the space filling
SD8-brane discussed earlier. In string frame the metric is written
\beq ds^{2}_{S8} = -dt^{2} + a(t)^{2}d{\vec y}^{2}. \eeq Upon
converting to Einstein frame we get the metric \beq ds_{ES8}^{2} =
-d\tau^{2} + a_{E}(\tau)^{2}d{\vec y}^{2}, \eeq where we have used
a change of coordinate such that $d\tau^{2}=e^{-\Phi/2}dt^{2}$,
and where \beq a_{E}(\tau) = e^{-\Phi(t)/4} a(t). \eeq For the
SD8-brane we found
\begin{eqnarray} \lim_{t\rightarrow \pm\infty} a(t) = 0 \, , \;\;\;\;
\lim_{t\rightarrow \pm\infty} e^{\Phi(t)} = 0.
\end{eqnarray} In string frame this implies that the metric
``closes~off'' at infinity but in the (more physical) Einstein
frame the converse happens, {\it i.e.}, the limit $\tau\rightarrow
\pm \infty$ corresponds to a constant scale factor, \beq
\lim_{t\rightarrow \pm \infty} a_{E}(\tau) = {\rm const.} \eeq
This is a very sensible result. We mentioned before that, because
of the absence of a transverse space, there appeared to be no
channel into which the energy could go. We therefore find that
asymptotically the energy has gone into inflating the worldvolume
of the SD-brane. In fact, the relaxation time for the
gravitational field is essentially infinite.

\subsection{Other classes of solutions} \label{othercases} In this section we
present a more general analysis of the family of solutions with
parameters $\{k_{\parallel},k_{\perp}\}$ associated with the metric
ansatz (\ref{genmet}).

\subsubsection{Comment about the numerical analysis} For
$\lambda=0$ there is clearly no source for the the bulk fields.
The equation of motion for the tachyon is then (\ref{theeom}). In
Appendix \ref{flatach} we found analytic expressions for the
corresponding solutions. They have the large time property \beq
\label{asympbe} \frac{V(T)}{\sqrt{\Delta}} = {\rm const.} \eeq
However, if we solve eq.~(\ref{theeom}) numerically (using the
same techniques as in section~\ref{section:numeric}) we find that
$V(T)/\sqrt{\Delta}$ appears to become constant but, after some
more time has elapsed, its behavior becomes erratic, {\it i.e.},
it starts oscillating with increasingly large amplitudes. Clearly
this is only an artifact of the numerical analysis. The tachyon
evolution is such that \beq \lim_{t\rightarrow \pm \infty} V(T) =
0, \;\;\; \lim_{t\rightarrow\pm \infty} \sqrt{\Delta} = 0. \eeq
The source of the problem can be traced down to the numerics
having difficulties to evaluate a $\frac{0}{0}$ division at large
times.

The same thing happens for $\lambda\neq 0$. The quantity that
becomes ambiguous for large time is then \beq \label{quantity}
\frac{e^{\Phi(t)} V(T)}{\sqrt{\Delta}}. \eeq Again, the problem is
associated with the fact that both $e^{\Phi(t)}\, V(T)$ and
$\sqrt{\Delta}$ are zero for large time. It is important to
resolve this ambiguity because the quantity (\ref{quantity})
directly feeds in the equations of motions for the bulk fields. We
are able to show numerically that (at least for the solutions
considered in this work) for large time \beq \frac{e^{\Phi(t)}
V(T)}{\sqrt{\Delta}} = 0. \eeq Similarly to the $\lambda=0$ case,
past some finite time the behavior of the function
(\ref{quantity}) becomes erratic. To obtain numerical solutions
representing the evolution for all times we did the following:
past the time value where (\ref{quantity}) becomes ill-behaved, we
solve the system of differential equations without a source, {\it
i.e.}, for \beq \frac{e^{\Phi(t)} V(T)}{\sqrt{\Delta}} = 0, \;\;\;
e^{\Phi(t)}V(T)\sqrt{\Delta} =0, \eeq and appropriate boundary
conditions. This introduces negligible errors.

\subsubsection{Time-reversal symmetric solutions}\label{trssblar}
We consider the time-reversal symmetric solutions, {\it i.e.},
those associated with bulk fields having vanishing
time-derivatives at $t=0$: \beq
\dot{a}(0)=\dot{R}(0)=\dot{C}(0)=\dot{\Phi}(0)=0. \eeq The
constraint equation (\ref{fconstraint}) at $t=0$ is then \beq
\frac{V(T)}{\sqrt{\Delta}} = \frac{1}{\lambda e^{\Phi}}
\left(p(p+1)\frac{k_{\parallel}}{a^{2}}
+(7-p)(8-p)\frac{k_{\perp}}{R^{2}} \right). \eeq The LHS being
positive-definite, the constraint can be satisfied if and only if
the metric ansatz contains a subspace of positive curvature. These
consist in five categories of solutions, {\it i.e.},
$\{k_{\parallel},k_{\perp}\}=\{0,1\},\{-1,1\},\{1,-1\},\{1,1\},\{1,0\}$.
We have performed a detailed analysis of these solutions for $p=4$
and $\dot{T}(0)\leq 1/10$. Typically, we find that when the
tachyon has reached a point of its evolution where $|V(T)|\simeq
0$, labelled $t_{c}$, the solutions develop a curvature
singularity. The behavior of the bulk fields is as follows. The
time derivative of the scale factor is such that \beq
\lim_{t\rightarrow t_{c}} \dot{a} = -\infty, \;\;\;
\lim_{t\rightarrow t_{c}} a(t) = 0, \eeq which corresponds to a
big-crunch singularity on the ($p+1$)-dimensional worldvolume. The
cases $\{k_{\parallel}=-1,k_{\perp}=1\}$,
$\{k_{\parallel}=0,k_{\perp}=1\}$ and
$\{k_{\parallel}=1,k_{\perp}=0\}$ are such that \beq
\lim_{t\rightarrow t_{c}} \dot{R}(t) = -\infty, \;\;\;
\lim_{t\rightarrow t_{c}} R(t) =0, \eeq corresponding to the
transverse spherical space collapsing to zero-size in finite time.
For $\{k_{\parallel}=1,k_{\perp}=1\}$ and
$\{k_{\parallel}=1,k_{\perp}=-1\}$, we find \beq
\lim_{t\rightarrow t_{c}} \dot{R}(t) = +\infty, \;\;\;
\lim_{t\rightarrow t_{c}} R(t) = +\infty. \eeq For $k_{\perp}=-1$,
$R(t)$ goes to infinity faster than $t$. All solutions are such
that \beq \lim_{t\rightarrow t_{c}} \dot{\Phi}(t) = -\infty. \eeq
Generically, the large time behavior of the R-R field is
well-behaved, {\it i.e.}, \beq \lim_{t\rightarrow t_{c}}
\dot{C}(t) \approx 0, \;\;\; \lim_{t\rightarrow t_{c}} C(t) =
C_{c},\eeq where $C_{c}$ is finite but typically many orders of
magnitudes larger than the constants $C_{\infty}$ associated with
the regular solutions presented earlier. The case
$\{k_{\parallel}=0,k_{\perp}=-1\}$ is special because then the
time derivative of the R-R field diverges as well at finite time.

We believe our conclusions to be unaltered for other values of
$p$. We found no evidence that the singularities are resolved when
the time-reversal symmetry is broken.

\subsubsection{More regular solutions} Another class of candidate
SD-brane solutions we have studied are those with
$k_{\parallel}=0$ and $k_{\perp}=0$. The results we present for
these solutions are generic, {\it i.e.}, they hold for all $p$ and
reasonable boundary conditions ({\it i.e.}, first derivatives not
too large). Firstly, the solutions are always asymmetric around
$t=0$ since it was shown before that, for consistency, at least
one of the bulk fields must have non-vanishing kinetic energy at
$t=0$. The level of asymmetry will be reduced by having smaller
derivatives of the bulk fields at $t=0$. The corresponding
solutions are regular and asymptotically flat. We find that \beq
\lim_{t\rightarrow \pm \infty} \dot{a}(t) = 0, \;\;\;
\lim_{t\rightarrow \pm \infty} a(t) = 0, \eeq and \beq
\lim_{t\rightarrow \pm \infty} \dot{R}(t) = 0, \;\;\;
\lim_{t\rightarrow \pm \infty} R(t) = {\rm const.} \eeq In
Einstein frame both scale factors asymptote to non-zero constants.
For the dilaton we obtain \beq \lim_{t\rightarrow \pm \infty}
\dot{\Phi}(t) = 0, \;\;\; \lim_{t\rightarrow \pm \infty} \Phi(t) =
-\infty, \eeq {\it i.e.}, the string coupling vanishes
asymptotically. Finally the R-R field behaves like \beq
\lim_{t\rightarrow \pm \infty} \dot{C}(t) = 0, \;\;\;
\lim_{t\rightarrow \pm \infty} C(t) = {\rm const.} \eeq We note
that the relaxation time for these solutions is many orders of
magnitude larger than for the $\{k_{\parallel}=0$,
$k_{\perp}=-1\}$ cases presented earlier.

The two remaining cases are $\{k_{\parallel},k_\perp\}$ equal to
$\{-1,-1\}$ and $\{-1,0\}$. We found evidence that the corresponding
solutions are regular and asymptotically flat.

%====================================================================+
\section{Discussion}\label{section:discussion}

Our primary motivation for this work was the general problem of
seeking mechanisms for resolution of singularities in spacetimes of
interest in string theory.  SD-brane supergravity spacetimes presented
in refs.~\cite{gutperle,gutperle2,KMP} create somewhat of a
supergravity {\em emergency} because they are not only singular but
nakedly so.

An important first step in the resolution program was made in
ref.~\cite{buchel}, where some effects of unstable brane probes in
these backgrounds were considered.  In particular, the probe
physics was also sick, and taking this analysis seriously led to
an even more dire assessment of the likelihood of singularity
resolution without resorting to the inclusion of massive string
modes in both the open and closed sectors. In our work, then, we
began by plumbing the depths of the probe approach. We found it to
be generally insufficient for our purposes; one reason is that the
probe approximation takes itself out of its regime of
self-consistency.

We then launched into an investigation of the physics of the
gravitational fields exerted by SD$p$-branes for general $p$ by
including backreaction. In order to get started on this problem,
we had to make the approximation of considering only the most
relevant open- and closed- string modes, with full gravitational
backreaction taken into account. The equations we derived are
highly nonlinear and couple brane with bulk, so did not lend
themselves to solution analytically.  We therefore resorted to
numerical techniques to search for solutions. Because of this
restriction, we had to use an ansatz which smeared the branes in
the transverse space; this allowed us to turn the equations into
ODE's and integrate them numerically. An essential step was to
begin the numerical integration
near the {\em top} of the potential hill, and then reconstruct
asymptopia, which we were able to do successfully. Generically, the
solutions are time-reversal {\it asymmetric}. We have shown that the
time-reversal symmetric solutions with the correct R-symmetry are
unavailable. Given our two-stage approximation (lowest modes, and
smeared ansatz), we found it rather satisfying that significant
progress in the resolution program is already found at this level. In
particular, our solutions for rolling tachyons backreacting on
spacetime are {\em completely nonsingular}, and our approximations
satisfy the fundamental property of self-consistency. We find these
conclusions suggestive of resolution of the SD-brane spacetime
singularity emergency.

It is however hard to know for sure whether our nonsingular results
will survive refinement.  Therefore, let us now make some specific
remarks about technical roadblocks we encountered which forced us to
make approximations, their physical consequences, and future outlook.

Section~\ref{section:prelim} was where we derived the coupled
tachyon-supergravity equations for a general brane distribution,
assuming Chern-Simons terms are turned off in a consistent
truncation. Our resulting equations are on the one hand remarkably
non-robust, and on the other quite robust. What we mean by
non-robustness is this: our ability to obtain nonsingular
evolution depends importantly on the structure of these equations
of motion. Signs are crucial, coefficients are crucial, and so is
the inclusion of Ramond-Ramond fields.
In other words, our nonsingular results are highly specific to the
field couplings arising from the low-energy approximation to
string theory.  Other ``S-branes'' arising from
``string-motivated'' actions will probably not possess similarly
nonsingular behavior. The positive type of robustness we refer to
is also a desirable property.  What we see manifestly is that the
precise form of the potentials $\{V(T),f(T)\}$ is not important,
apart from the large-$|T|$ behavior which had been derived
elsewhere.  The most obvious refinement of our work here will be
to attack the problem of relaxing the requirement of zero NS-NS
$B$-field.  Allowing $B^{(2)}$ to be turned on will allow us to
break $ISO(p{+}1)$ on the worldvolume and allow inhomogeneous
tachyonic modes --- the importance of which is discussed in
refs.~\cite{cosmotachyonK,larsen} --- and also to turn on more
components of R-R fields.  Inhomogeneities would of course have to
be included in initial conditions, because homogeneous on-shell
tachyons do not couple to non-homogeneous tachyons
\cite{kutasov2}.
We have postponed the non-homogeneous problem to the future mainly
because it is messy; our work reported here should be considered a
step in a larger program.

The other important approximation we made in our work was in
section~\ref{subsection:specific_ansatz}, where we had to smear
the SD-brane sources in the transverse space to facilitate
integrating the equations numerically by turning them into ODE's.
This limits our ability to fully probe the properties of the
system in which we are interested.  Here we would also like to
record another physical consequence of this ansatz.  Namely, this
restriction has notable, negative, consequences for our ability to
track whether black holes form as intermediate states during the
time evolution of our coupled system including full backreaction.
The issue of black holes was raised in the discussion section of
ref.~\cite{KMP}.  The essential point is that a black hole
intermediate state may arise as an alternative to SD-brane
formation and decay, at least with the half-advanced,
half-retarded propagator.  The fine-tuned nature of the initial
conditions producing SD-branes highlights a reason why integrating
partial differential equations of motion (including dependence on
transverse coordinates) may be particularly difficult numerically.
Or the obstruction to finding the full solution may yet turn out
to be negotiable. It will also be interesting to think further
about particle/string production.

Let us end with some somewhat speculative remarks.  Typically in
the limit $g_{s} N\rightarrow 0$ the open and closed strings
decouple. This is true in our effective lowest-modes analysis
here, but also explicit in other worldsheet-inspired approaches.
There should also exist a limit (in time) to be taken where only
the open strings survive. From the worldsheet definition of a
SD-brane, it is suggestive that the open string degrees of freedom
would combine to form a Euclidean conformal field theory in $p+1$
dimensions. The same appears true when considering the effective
action of massless open string degrees of freedom on an unstable
D-brane \cite{hashimoto}. In both approaches, however, it is not
clear what the role of the tachyon could be. Physically, it is the
source of a process by which energy is siphoned out of the open
string sector and pumped into the closed string sector. So, in a
sense, the decay of a D-brane through tachyon condensation
corresponds to the decrease of a $c$-function-like quantity on the
gauge theory side. Then, we can entertain the idea that
time-evolution on the gravity side should really be regarded as a
renormalization group (RG) flow on the gauge theory side. From
this viewpoint, formation and decay of a SD-brane would be a
process corresponding to first an {\em inverse} RG flow
(integrating in degrees of freedom) followed by regular RG flow
(integrating out degrees of freedom).\footnote{Similar ideas were
explored in the context of the dS/CFT correspondence (see, for
example, refs.~\cite{andycft,us})} This might be related to the
study of open string tachyon condensation using RG flow in the
worldsheet theory \cite{martinec}.

%--------------------------------------------------------------------+
\vskip0.1\textheight
%--------------------------------------------------------------------+
\section*{Acknowledgements}

The authors wish to thank Alex Buchel, Lev Kofman, Martin
Kruczenski, Finn Larsen, Juan Maldacena, Alex Maloney, Don Marolf,
Rob Myers, Andy Strominger, and Johannes Walcher for useful
discussions. Finally we thank Dave Winters for proof-reading an
earlier draft of this paper.

FL was supported in part by NSERC of Canada and FCAR du Qu\'ebec.  AWP
thanks the Radcliffe Institute for Advanced Study, and the High-Energy
Theory group of the Harvard University Physics Department, for
hospitality during Fall semester 2002-3 while this work was carried
out.  Research of AWP is supported by the Radcliffe Institute, CIAR
and NSERC of Canada, and the Sloan Foundation of the USA.

%====================================================================+
\newpage
%\vskip0.1\textheight
%--------------------------------------------------------------------+
\appendix
%====================================================================+
\section{Regular KMP SD-brane solutions} \label{regular}

Among the supergravity solutions found by \cite{KMP}, there exist some
that are regular on the horizon located at $t=\omega$. This is only
realized for the following values of the parameters, \beq \tilde{k} =
2, \;\;\; H = \frac{4}{7-p}, \;\;\; G=k_{i}=0. \eeq The corresponding
metric is
\begin{eqnarray} ds^{2} = && F(t)^{1/2}\alpha(t)^{4/(7-p)}
\left(-dt^{2}+t^{2}dH_{8-p}^{2}\right) \nonumber\\
&&+ F(t)^{-1/2}\left[\sum_{i=2}^{p+1}\left(dx^{i}\right)^{2}
+\left(\frac{\beta(t)}{\alpha(t)}\right)^{2}\left(dx^{1}\right)^{2}
\right].\end{eqnarray}
Because these solutions are anisotropic in the worldvolume
directions, it is not clear that they are physically relevant. We
will nevertheless study some interesting properties not considered
in ref.~\cite{KMP}. For example, the region $t=+\omega$ does {\it
not} correspond to an horizon as suggested by the fact that none
of the metric components either vanish or diverge there. For $p$
{\it odd} the solutions are time-reversal symmetric so the region
$t=-\omega$ is also not an horizon.

%--------------------------------------------------------------------+
\subsection{The region close to the origin}

For the anisotropic solutions there is no curvature singularity at
$t=\omega$. It is therefore interesting to consider the behavior
of the metric components and curvature invariants close to the
potentially problematic region around the origin, $t=0$. We
evaluated the curvature invariants: ${\cal R}$, ${\cal
R}_{\mu\nu}{\cal R}^{\mu\nu}$, and ${\cal
R}_{\mu\nu\rho\lambda}{\cal R}^{\mu\nu\rho\lambda}$. They
identically vanish for $t=0$. The metric tensor there appears
suspicious (for example, the component $g_{tt}$ diverges) but we
find \beq \label{limit} \lim_{t\rightarrow 0} ds^{2} = - d\tau^{2}
+ \tau^{2}dH_{8-p}^{2} + \sum_{i=1}^{p+1}\left(dx^{i}\right)^{2},
\eeq where $\tau=\omega^{2}/t$. The expression (\ref{limit}) is
simply flat space with part of it written in Milne coordinates.
Not surprisingly, we find \beq \lim_{t\rightarrow 0} \dot{\Phi}(t)
= 0, \;\;\;\; \lim_{t\rightarrow 0} \dot{C}(t) = 0, \eeq which
implies that all stress-energy components vanish in the the region
close to the origin.

%--------------------------------------------------------------------+
\subsection{Horizon physics}

We demonstrate that, contrary to previous claims, many of the
anisotropic solutions are actually regular in the {\em full}
range: $-\infty<t<+\infty$.  The anisotropic solutions were
already shown to be non-singular at $t=\omega$ and $t=0$. We now
investigate the region $t=-\omega$ further. Let us introduce the
change of coordinates \beq T = \left( 1 +
\left(\frac{\omega}{t}\right)^{7-p} \right)^{2/(7-p)}t \eeq in
order to make comparison with the results of ref.~\cite{KMP}
easier. For $p$ even, $T=0$ corresponds to $t=-\omega$ while for
$p$ odd we have $T=-2\omega$ when $t=-\omega$. A comment in
ref.~\cite{KMP} is that $T=0$ corresponds to a (non-naked)
curvature singularity. Actually, this is not always the case! For
example, we considered the case $p=1$ and computed the associated
curvature invariants at all times. Figure \ref{fig1} shows the
evolution of the Ricci scalar for the solution with $\omega=1$ and
$\theta=\pi/4$. Clearly, the $p=1$ solution is symmetric under
time-reversal and therefore has no curvature singularity. It also
does not possess any horizon, a feature common to the regular
solutions found in this paper. The qualitative behavior of all
curvature invariants is similar to the Ricci scalar and is quite
generic, {\it i.e.}, it is unchanged for all {\it odd} values of
$p$, $\theta$ and $\omega$. For $p=1$ we obtain \beq
\lim_{t\rightarrow \pm \omega} {\cal R} = -\frac{3
2^{1/3}}{\omega^{2}}\frac{19\cos^{4}\theta -26\cos^{2}\theta +
7}{\sin^{5}\theta}. \eeq This is finite except for $\theta=0$ in
which case the Ricci scalar diverges like $R\sim
1/(t-\omega)^{3}$. We also found expressions for two other
curvature invariants (for $p=1$), \beq \lim_{t\rightarrow \pm
\omega} {\cal R}_{\mu\nu}{\cal R}^{\mu\nu} =\frac{9
2^{3/4}}{2\omega^{4}} \frac{-272 \cos^{6}\theta + 14 + 255
\cos^{4}\theta + 101\cos^{8}\theta
-98\cos^{2}\theta}{\sin^{10}\theta}, \eeq \beq \lim_{t\rightarrow
\pm \omega} {\cal R}_{\mu\nu\rho\sigma}{\cal R}^{\mu\nu\rho\sigma}
= \frac{3 2^{2/3}}{4 \omega^{4}}\frac{-112\cos \theta + 28 +892
\cos^{8}\theta -1840 \cos^{6}\theta +1032 \cos^{8}\theta
}{\sin^{10}\theta}. \eeq We found expressions with similar
qualitative behavior for other values of $p$ {\it odd}.

For $p$ {\it even} the solutions are not time-reversal symmetric.
As pointed out previously, curvature invariants are finite at
$t=\omega$ but there is a curvature singularity at $t=-\omega$.
These are the spacelike curvature singularities (protected by an
horizon at $t=0$) described in ref.~\cite{KMP}.

\FIGURE{\epsfig{file=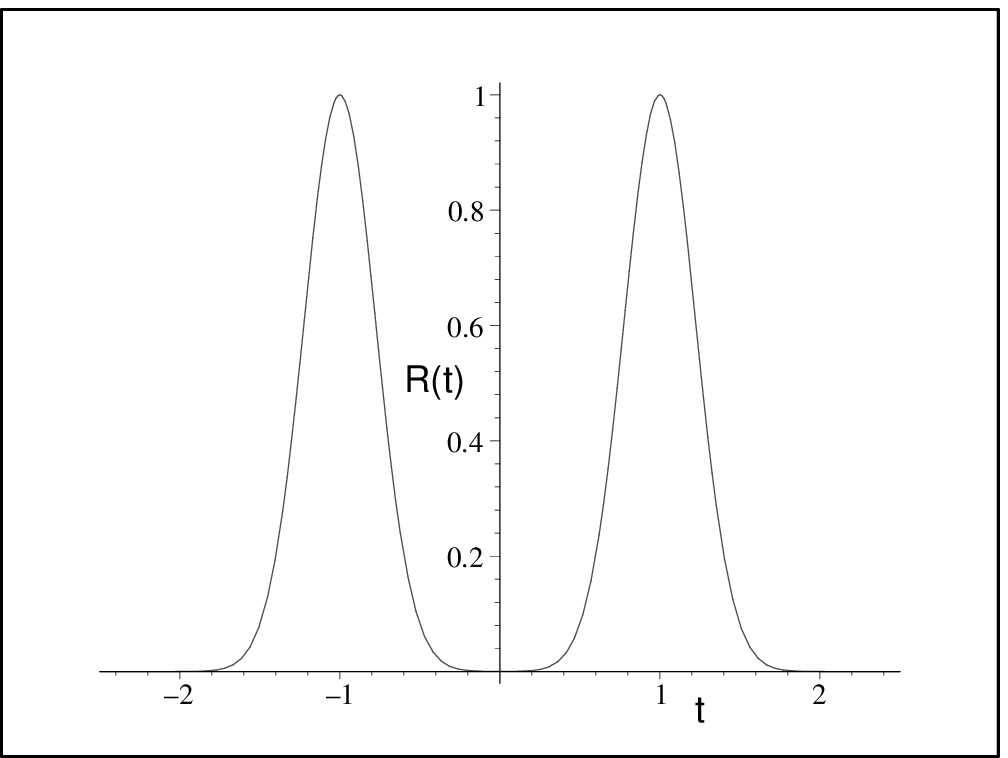}\caption{This figure illustrates
the Ricci scalar for the $p=1$ anisotropic SD-brane solution with
$\omega=1$ and $\theta=\pi/4$. The other curvature invariants
behave similarly.} \label{fig1}}

%--------------------------------------------------------------------+
\subsection{Unstable brane probe analysis}

As mentioned in section \ref{subsection:buchel}, the motivation
behind considering an unstable brane probe in a background with
singularity problems is to ask if the singular background could
actually be built.

The calculations and results of ref.~\cite{buchel} were summarized
in section \ref{subsection:buchel}. In this appendix we generalize
this calculation by probing the $d=10$ {\em anisotropic}
backgrounds presented above, since these are the only ones which
are either non-singular or have singularities (at $t=-\omega$)
shielded by a horizon ($t=0$). We felt this generalization to be
necessary because ref.~\cite{buchel} did not, for example, address
the issue as to how the inclusion of the dilaton might affect the
brane probe calculation.  The unstable brane action is the obvious
generalization eq.~(\ref{sourcebuc}) of the case studied in
ref.~\cite{buchel}.

We investigate whether or not an unstable brane probe is a
well-defined object in the vicinity of the region $t=\omega$. In
Einstein frame, the energy density for the probe propagating in
the anisotropic backgrounds is \beq \rho_{probe} =\frac{N
\mu_{p+1}}{g_{s}} f(\Phi) \frac{V(T)}{\Delta^{1/2}}, \eeq while
the pressure corresponds to \beq p_{probe} =\frac{N
\mu_{p+1}}{g_{s}} f(\Phi) V(T) \Delta^{1/2}. \eeq The dilaton
function $f(\Phi)$ was picked up during the transformation from
the string frame to the Einstein frame. It plays no role in the
upcoming analysis because the dilaton is well-behaved, \beq
\lim_{t\rightarrow \omega} f(\Phi)= {\rm const.} \eeq

As we saw in section \ref{subsection:buchel}, whenever the probe
analysis goes wrong, it signals a pathology for the gravitational
background.  As we saw, there are at least two ways the probe
analysis can go wrong: it may induce an infinite energy or
pressure density ($\rho_{probe}, p_{probe} \rightarrow \pm
\infty$), or, there might not exist any reasonable solutions for
$T(t)$.

For $t\simeq \omega$, the dominant contribution to the equation of
motion for the tachyon is \beq \Delta^{2} - \Delta + \frac{2}{9}
(t-\omega) \dot{\Delta} = 0. \eeq This is solved for \beq
\Delta(t) = \frac{(t-\omega)^{9/2}}{(t-\omega)^{9/2}-g},\eeq where
$g$ is a constant of integration. The solution $\Delta=0$ ($g \neq
0$) clearly corresponds to the brane probe inducing a curvature
singularity on the horizon. The only physical solution is the one
for which $g=0$ which corresponds to $\Delta=1$. For the
anisotropic backgrounds considered here the metric component
$g_{tt}$ neither vanishes nor blows up at $t=\omega$. The solution
$\Delta=1$ therefore corresponds to a tachyon field for which the
time-derivative vanishes ($\dot{T}=0$) at $t=\omega$. It therefore
appears that there are solutions for the probe evolution that
avoids the pathologies described earlier. This is no surprise
since for these anisotropic backgrounds the region $t=\omega$ is
not an horizon in the technical sense of the term. We repeated the
calculation around the regions $t=0$ and $t=-\omega$. We find that
for both $p$ {\it odd} and {\it even} the unstable brane probe is
not well-behaved at $t=0$, {\it i.e.}, it induces a curvature
singularity there.

%====================================================================+
\section{Tachyon in flat space}\label{flatach}

We consider solutions to the equation of motion for an open string
tachyon when the massless closed string modes are decoupled. The
relevant equation of motion is \beq \label{eomt_two} \ddot{T} +
(1-\dot{T}^{2})\frac{\partial \ln V(T)}{\partial T} = 0. \eeq

\subsection{General solution} Eq.~(\ref{eomt_two}) is a second
order differential equation with a general solution of the form
\beq T(t) = \int dt \; \frac{1 + a^{2}\, V^{2}(t)}{1-a^{2}\,
V^{2}(t)} + b, \eeq where $a$ and $b$ are constants of
integration. To integrate this equation one needs the function
$V(t)$ which would imply that we already know the solution $T(t)$.
Open string field theory has taught us that for $t=t_{c}$ large we
have \beq \lim_{t \rightarrow t_{c}} V(t) << 1, \eeq which implies
that \beq T(t) \simeq \int_{t_{c}}^{t} dt \; (1 + 2 a^{2} V(t)) +
b. \eeq Therefore at large time the tachyon behaves like \beq T(t)
= t + (b-t_{c}) + 2a^{2} \int_{t_{c}}^{t} dt \; V^{2}(t). \eeq
Using the string field theory result \beq \lim_{T\rightarrow
+\infty} V(T) = e^{-T/\sqrt{2}}, \eeq and taking $T(t)\simeq t$
leads to the large time formula \beq T(t) \simeq t -a^{2}\sqrt{2}
e^{-\sqrt{2}t}, \eeq where we have fixed the integration constants
by imposing \beq a^{2}\sqrt{2} e^{-\sqrt{2}t_{c}} + b - t_{c} =
0.\eeq

\subsection{Particular solutions} We consider solutions to the
equation of motion (\ref{eomt_two}) with the potential \beq V(T) =
\frac{1}{\cosh \left(T/\sqrt{2}\right)} \, . \eeq The tachyon
equation of motion becomes \beq \label{theeom} \ddot{T} +
\frac{1}{\sqrt{2}}(1-\dot{T}^{2}) \tanh \left(T/\sqrt{2}\right),
\eeq which has a solution of the form \beq T(t) = -\sqrt{2}\; {\rm
arc\, sinh}\; \left( \frac{\sqrt{2}}{2} \left[ c_{1}
e^{t/\sqrt{2}} - c_{2} e^{-t/\sqrt{2}} \right] \right), \eeq where
$c_{1}$ and $c_{2}$ are constants of integration. We usually
specify boundary conditions at $t=0$, \beq T(0)=-\sqrt{2} \;{\rm
arc\, sinh} \; \left(\frac{\sqrt{2}(c_{1}-c_{2})}{2}\right), \eeq
\beq \dot{T}(0)=-\sqrt{2}
\frac{c_{1}+c_{2}}{\left(4+2(c_{1}-c_{2})^{2}\right)^{1/2}}.  \eeq
The family of solutions characterized by $T(0)=0$ ($c_{1}=c_{2}$)
corresponds to all possible tachyon velocities at $t=0$:
$\dot{T}(0)=-\sqrt{2} c_{1}$. An other class of solutions are
those for which $\dot{T}(0)=0$ ($c_{1}=-c_{2}$). Those correspond
to allowing the tachyon to begin its evolution with
$T(0)=-\sqrt{2}\;{\rm arcsinh} \; \sqrt{2}c_{1}$.

The solution we presented are referred to as {\it tachyon matter}.
The stress-energy components (which are independent of the number
of dimensions in the theory) correspond to a conserved energy
density ($\rho \sim V(T)/\sqrt{\Delta} = {\rm constant}$) and a
pressure ($p\sim -V(T)\sqrt{\Delta}$) that vanishes as
$t\rightarrow +\infty$ \cite{sen2}.

%====================================================================+

%====================================================================+
\end{document}